**NUMBER 116   AUGUST 2009**

# AAO NEWSLETTER

**ANGLO-AUSTRALIAN OBSERVATORY**

## AAOmega maps the ISM towards ω Centauri

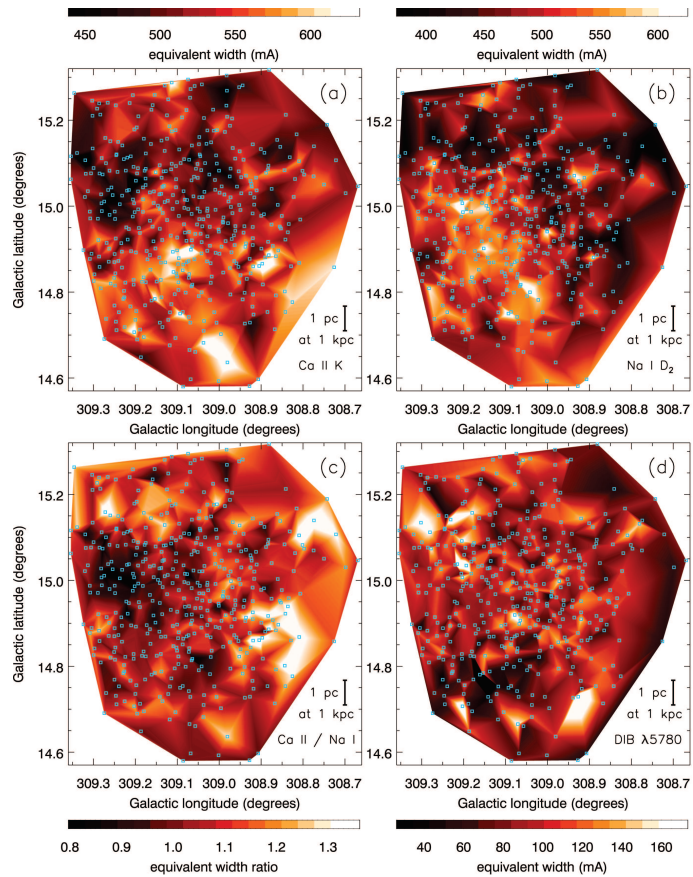

Equivalent width maps of the [Ca II] K line, [Na I] $D_2$ line, their ratio, and the Diffuse Interstellar Band at 5780 Å, in Galactic coordinates. The strength of these tracers relative to one another depends on the conditions in the gas. Structure is present both on small and large scales, across the region and along the path through the Galactic Disc and Halo.

# DIRECTOR'S MESSAGE

I'm delighted to report that the AAO's future beyond the end of the AAT Agreement on 30 June 2010 has now been secured. With the final withdrawal of the UK on that date, the Anglo-Australian Observatory will become the Australian Astronomical Observatory, operating under the Department of Innovation, Industry, Science & Research (DIISR). Funding for the final year of the old AAO (2009-10) and the first three years of the new AAO (2010-13) was provided in the Australian Government Budget released in May. The funding for the new AAO will allow the Observatory to continue to provide the level of services and support that it currently maintains for Australian users of the AAT, Gemini and Magellan, together with its world-renowned programs of astronomical research and instrumentation. This brings to an end several years of uncertainty over the future of the organization and provides a degree of funding certainty that will allow the AAO to plan more effectively for the future.

The AAO's executive team are now working on the process of transition to the new organizational format. This involves dealing with a large number of issues ranging from the legislation needed to effect the change through to ensuring that the AAO's procedures and practices mesh with those of DIISR. One of our goals is to make this process as transparent as possible both to AAO staff and to users of the AAO's telescopes and other facilities. Of course, for UK users there will no longer be a UK share in the AAT, however all international users will continue to have substantial access to the telescope under a policy that the AAT Board will be considering at their upcoming meeting in September.

Two current projects that are also shaping the AAT's future are the major NCRIS-funded efforts to refurbish the AAT and construct the HERMES spectrograph. The AAT refurbishment project is proceeding well, with replacement and upgrading of several major telescope infrastructures, including the dome and telescope axis encoders, the fire alarm system, the primary mirror elevator, and the air-conditioning and ventilation system. Additional measures have also been taken to further improve dome safety for staff and users. The HERMES project is also making good progress. The conceptual design phase took longer than expected, but has resulted in a powerful and versatile instrument. The baseline design for HERMES is a three-channel spectrograph that utilizes VPH gratings for maximum efficiency. A fourth arm can optionally be implemented if funds are available, and would significantly enhance the power of HERMES to carry out stellar chemical abundance surveys of the Milky Way, its primary science mission. The Preliminary Design Review for HERMES is scheduled for 30 September.

Other projects either underway or seeking funding include the CYCLOPS fibre feed for UCLES, expected to be commissioned and available for shared-risk observing in Semester 2010A; an OH-suppression fibre feed system called GNOSIS, initially intended for IRIS2 on the AAT and potentially thereafter for GNIRS on Gemini; and the NG1dF concept for a prime focus multislit spectrograph with a 1-degree field of view on the AAT.

The funding for the new AAO is only part of a wider picture of strong Australian Government support for astronomy that also includes additional funding for the Australian SKA Pathfinder (ASKAP) and an SKA science and data centre, the naming of Space Science & Astronomy as one of three 'Super Science' initiatives that will include 30-40 new early-career research fellowships in the field, and the provision of $88.4M to pay for a 10% share in the construction of the Giant Magellan Telescope (GMT) and enhance Australian involvement in the telescope and instrument contracts. Part of this funding will support the AAO's proposal to carry out a concept study for a facility fibre feed system for GMT called MANIFEST.

It is wonderful to be able to report so much good news for the future of the AAO and Australian astronomy in general!

Matthew Colless

# IRIS2 OBSERVES KAGUYA'S DEMISE

Jeremy Bailey (UNSW), Steve Lee (AAO) & Hakan Svedhem (ESA/ESTEC)

On the morning of June 11 (local time) the Japanese lunar spacecraft Kaguya was purposely crashed into the moon, thus terminating its survey mission in a very spectacular fashion. The resulting impact was successfully observed with IRIS2 on the Anglo-Australian Telescope.

IRIS2 was used in time series mode with 1 second exposures (with 0.6 seconds readout time between each frame). The bright moon (just past full) was placed as far to one side of the frame as possible, while a 2.3 micron narrow band filter was used to keep the remaining light from saturating the detector too badly. A bright impact flash was seen within a few seconds of the predicted time, at 18:25:10 UT June 10. The four frames here from around impact time show the bright flash in the second frame, and also faintly visible in the third and fourth.

There was only one other successful observation of this impact despite a wide campaign to observe it from a number of locations. Both observations were made at infrared wavelengths. Together with previous observations of the impact flashes of SMART-1 (from the CFHT) and of HITEN (from the AAT) it now seems clear that these bright impact flashes are only seen at infrared wavelengths and are not seen in the visible.

The value of such events lies in the fact that the properties of the impactor (mass and velocity) are accurately known and hence we can link the impactor properties to observations of what happened during the event, and eventually to the properties of the crater produced (which should be observable from future missions). Such observations thus help us to understand the impact process, a process that is extremely important in shaping the surfaces of many solar system bodies.

Of particular interest is to investigate how well a meteorite might survive an impact on the Moon. The Kaguya impact occurred at a very shallow angle, which maximises the chance of partial survival. Meteorites from Earth that impacted on the Moon early in its history (during the late heavy bombardment) might preserve a record of Earth's early geological history that is largely lost on Earth (Crawford et al., 2008, Astrobiology, 8, 242–252).

NASA has recently launched a spacecraft called LCROSS, which is designed to impact in the Moon's polar regions, where it is suggested that ice might exist inside permanently shaded craters. Here it is hoped that observations of the impact would enable the presence of ice to be detected. This ice could be a valuable resource for a future lunar base.

We would like to thank the Director and scheduler for allocating time for this observation at very short notice.

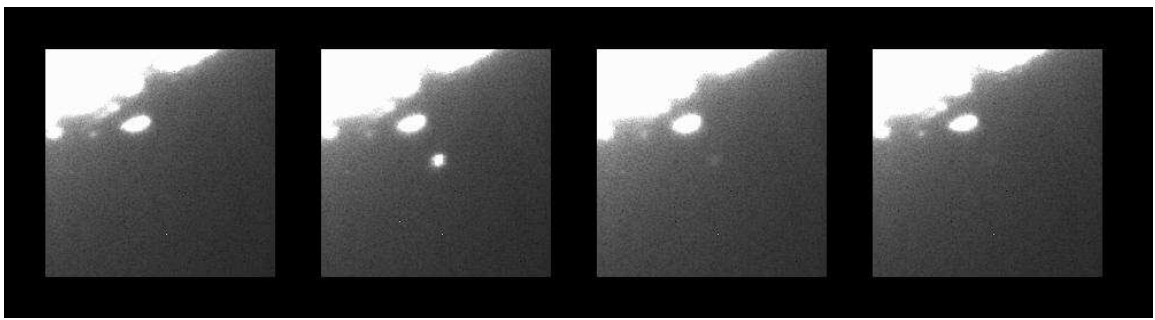

Photo courtesy of University of New South Wales/Anglo-Australian Observatory (J. Bailey and S. Lee)

# HOW THE BRIGHTEST GLOBULAR CLUSTER SHINES A LIGHT ON DIFFUSE GAS IN OUR GALAXY

Jacco Th. van Loon (Keele), Keith T. Smith (Nottingham), Iain McDonald (Keele), Peter J. Sarre (Nottingham), Stephen J. Fossey (UCL), Robert G. Sharp (AAO)

## Abstract

Exploiting the unsurpassed multiplexity and superb throughput of the AAOmega spectrograph on the AAT, the first absorption-line maps were constructed of the interstellar medium in front of the largest and most glorious Galactic globular cluster, $\omega$ Centauri. Lines of neutral sodium, singly-ionised calcium and diffuse interstellar bands, in combination with dust reddening maps and an analytical model for an ensemble of clouds, were used to build a self-consistent picture of the different components and density and ionisation fluctuations therein. Extra-planar gas is demonstrably structured at parsec scales, dominated by a harsh radiation field and possibly lagging behind Galactic rotation.

## A wealth of detail in the darkness of space

Interstellar space is a better vacuum than any created in laboratories on Earth. But that is not to say it is empty! The diffuse matter of the interstellar medium (ISM) makes itself known most conspicuously in regions of recent star formation, in the form of reflection nebulae and recombination radiation from hydrogen and oxygen, both due to the young massive stars illuminating the surrounding dust and gas. The Infrared Astronomical Satellite in the 1980s produced stunning images of the glow from interstellar cirrus, the thin dust filaments veiling our views through the Galactic Disc. Tuning in at the radio frequency of the spin-flip in the hydrogen atom, it becomes evident that gas is around everywhere in the Milky Way. Indeed, gas clouds even populate the Galactic Halo, far from the spiral arms that give our Galaxy its characteristic appearance. Absorption of background light by highly-ionised atoms is seen by spacecraft in any direction, even in the voids between interstellar emission.

The diverse collection of observations amalgamated into the paradigm of the multi-phased ISM, where hot gas (up to millions K) occupies a large volume and cold gas (down to 15 K) defines the smaller, denser clouds. These different phases are determined by the cooling and heating functions, which in turn depend on the phase of the medium as well as on the density, on the radiation field, and on shock waves that travel through the ISM even without the need for explosions or jets. The result is a myriad of intricate structures, from bubbles and shells to cometary clumps, sheets and strings. The scales range from Galactic down to that of the Solar System – even in the warm, weakly-ionised medium. The interfaces that separate different phases generally do not seem to be in pressure-equilibrium (although the contribution of magnetic fields to the total pressure, in addition to the thermal and turbulent motions, and its acting to confine the flow of charged fluids are uncomfortably ill-constrained). Even if they were, the ISM is far from a static ensemble of clouds but rather a highly dynamic place where morphological and thermo-dynamical structure is as evanescent as it is elusive.

## Globular clusters as pencils drawing a painting of the sky

Among the earliest recognitions that interstellar space is not empty, the spectral appearance of a star is affected by scattering and absorption by matter along the sightline. Dust grains dim the star, more so at blue than at red wavelengths, causing interstellar reddening. Some atoms and molecules have strong electronic transitions at optical frequencies, which cause sharp absorption lines in the stellar spectrum, displaced from the photospheric lines. The most mysterious of all, the Diffuse Interstellar Bands (DIBs) have been known for over a century but their carriers remain unknown. All these signatures trace the cold and warm neutral medium, and the warm ionised medium. Their column densities and kinematics can be used to map structures, by using many stars or by following the signatures in a star's spectrum as it moves through space. Observing multiple signatures may allow one to better constrain the ionisation and excitation conditions, which otherwise can mimic fluctuations in gas density. The absorption-line studies can probe the ISM that is too feeble to image directly.

Globular clusters are great objects to use to probe the intervening ISM. The massive ones contain millions of stars, thousands of which are bright and not too littered with spectral lines of their own. As they are all at practically the same distance we can be sure that differences in the interstellar signatures reflect transversal structure. Depending on the distance and compactness of the cluster, and whether one looks toward the central portions of the cluster or its periphery, distances between the stellar probes can range from an arcsecond to nearly a degree on the sky, translating to anything between thousands of astronomical units and dozens of parsecs. Of course, one is restricted to the directions toward the globular clusters.

In the 1990s, stars in several globular clusters were

SCIENCE HIGHLIGHTS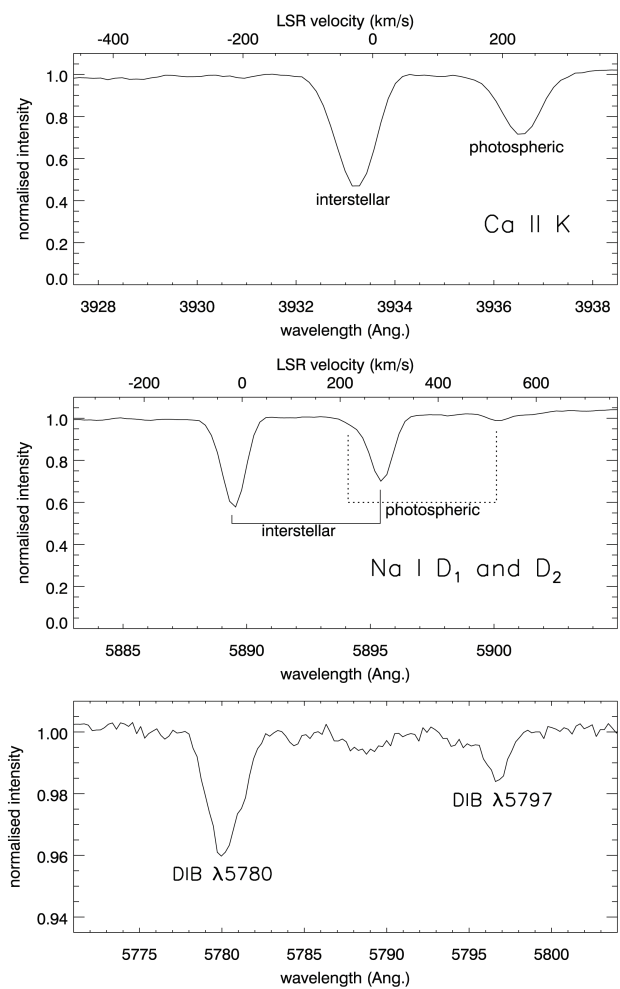

Figure 1: Average spectra of the [Ca II] K line, [Na I] $D_1$ and $D_2$ lines, and Diffuse Interstellar Bands at 5780 and 5797 Å. The low metal content of stars in ω Centauri results in a clean photospheric continuum, against which the interstellar lines are seen clearly, and well-separated from the photospheric components thanks to the high space motion of ω Centauri.

targeted for the purpose of ISM studies. The signatures used were almost exclusively the calcium H and K and sodium D (1 and 2) Fraunhofer lines near 390 and 590 nm, respectively. Although these are definitely not the most abundant metals their lines are strong and fall within the optical regime. These studies demonstrated that structure exists in the warm neutral and ionised medium down to the resolution limit, on sub-parsec scales. However, as they only comprise a small number of stars per cluster these structures could not be mapped in their full splendour.

**AAOmega on ω Centauri**

The most massive Galactic globular cluster, ω Centauri is also one of the nearest. Thus it subtends a vast area on the sky, about a degree in diameter, and its stellar members are bright and well separated. As it lacks the central cusp typical of core-collapsed clusters even the core of ω Centauri is accessible. But the cluster is highly suitable for ISM studies for other reasons, too. It has a well-populated Blue Horizontal Branch (BHB) of hot (8 000 to 30 000 K) metal-poor stars (most stars in ω Centauri have about 50 times fewer metals than the Sun, though a minority have up to ten times as many). These stars have very weak absorption lines, except for hydrogen and helium in the hottest examples, making it much easier to spot the interstellar components. As if this is not sufficient, ω Centauri moves on a retrograde orbit, displacing its absorption lines by more than 200 km/s from the Galactic ISM. We should be so lucky!

In the year 2000, we pointed the 2dF spectrograph on the AAT at over 1500 post-main sequence stars that had just been confirmed as proper-motion members of ω Centauri (van Loon et al. 2007). Although intended mainly for constructing a roadmap of the post-main sequence evolution of metal-poor low-mass stars, we could barely use the calcium K line to construct a map

ANGLO-AUSTRALIAN OBSERVATORY
NEWSLETTER
AUGUST 09

page 5

of the interstellar absorption across the face of the cluster. Forced to smooth heavily, the only feature that could be made out was an enhanced column density of absorbers somewhat offset from the cluster centre. To do this better would require higher spectral resolution whilst preserving the low noise level.

Enter AAOmega, the successor to the 2dF spectrograph. Using the same robotic fibre positioner, but with much improved throughput and finer gratings, spectra could be obtained of hundreds of stars at a good enough signal and a resolving power $\lambda/\Delta\lambda = 8\,000$, four times superior to the old 2dF spectrograph. With only two fields, using just six hours of AAT time in service mode early last year, spectra were obtained of 452 BHB stars (van Loon et al. 2009), more than an order of magnitude more ISM probes than in any previous globular cluster study. This allowed us for the first time to construct maps of roughly 20 by 20 resolution elements, sparsely but fairly uniformly sampled. The ISM signatures covered were principally the calcium H and K and sodium D lines, but in a large number of stars the strongest DIBS, at 5780 and 5797 Å, were detected.

**Small-scale structure in the multi-phased Disc-Halo interface**

Our maps show structure on scales of tens of arcminutes, which correlates weakly with the distribution of dust depicted in the IRAS/DIRBE reddening map. This suggests that we probe different neutral clouds, by their increments to column density. However, the calcium and sodium maps differ. The neutral sodium traces predominantly the neutral medium whereas the singly-ionised calcium traces both the warm neutral and the warm ionised medium. We also detected the effect of depletion of calcium from the gas phase by its incorporation into dust grains. Given the relatively low interstellar visual extinction of about 0.1 magnitude at this Galactic latitude of 15 degrees, the moderate level of depletion is as expected.

One might be deluded into thinking that such unobtrusive sightlines would be simple, boring even. No such thing! With the benefit of hindsight of literature studies at high spectral resolution along a few sightlines, and consistent with our kinematic maps, the sodium maps appear to be dominated by material above the Carina-Sagittarius spiral arm at 1–1.5 kpc distance, whereas the calcium maps contain a significant component arising in the extra-planar inner-Halo region situated beyond the spiral arm. The behaviour of the DIBs is consistent with this interpretation, and suggests that clouds in the Disc-Halo interface are predominantly of the type where the 5780 Å DIB is relatively strong – this is taken as a sign of a strong radiation field. Curiously, comparison with an analytical absorption-line profile suggests that extra-planar gas may be lagging behind the Galactic rotation curve, possibly as a result of friction between the Disc and Halo. One may speculate that this could incite instabilities which propagate throughout the Disc giving rise to the spiral arm pattern.

The fluctuations seen on scales of arcminutes are due to a combination of fluctuation in density and ionisation conditions. The former are well-reproduced with a simple analytical hierarchical cloud model, whilst the latter are inferred from the fact that reddening maps constructed from the coarse IRAS/DIRBE maps do a better job in tightening the various post-main sequence features in the colour-magnitude diagram than the finer maps derived from the premise that the sodium absorption traces reddening. In conclusion, our observations demonstrate that small-scale density and/or ionisation structures exist in the extra-planar gas.

**Next steps on the path of discovery**

There are several ways in which the present studies can be elaborated. There are an equal number of suitable BHB stars available, to add sightlines and thus resolution and fidelity to the ISM maps. These stars are more oddly placed and thus would require more than another two 2dF fields, but that is merely a slight inefficiency requiring a somewhat larger time allocation. The initial idea was that by re-observation of already observed targets, very much smaller scales would become available, given the known proper motions of the individual stars of several milli-arcseconds per year. A few years' delay would sample scales of about a hundred astronomical units. To be fair, the observations carried out with AAOmega were impressive yet revealed their limitations in terms of noise level and kinematic resolution, and it is doubtful that re-observation will yield fluctuations of an interstellar rather than a statistical nature. More highly spectrally-resolved measurements at a reduced noise level would be most worthwhile, and allow the proper-motion scanning technique to be applied. An upgrade of AAOmega with an even finer grating, in combination with a week's use of AAT time, could make this a reality.

# PUSHING THE DOPPLER LIMIT: FIRST RESULTS FROM THE ANGLO-AUSTRALIAN ROCKY PLANET SEARCH

Simon O'Toole (AAO), Hugh Jones (Hertfordshire), Chris Tinney, Rob Wittenmyer, Jeremy Bailey (UNSW), Paul Butler (Carnegie), Brad Carter (Southern Queensland)

Since the first detection of planets orbiting another Sun-like star in 1995, there has been a profound shift in astrophysics toward the planetary sciences. Research on extra-solar planets and the possibility of life in the Universe, which was not long ago considered to be at the fringes, is now at the heart of mainstream science, and the detection of Jupiter-like planets around other stars is now routine. The next challenges in exoplanetary research are to discover other Solar Systems like our own and to measure the underlying distributions of exoplanet properties to assess models of planetary formation and evolution.

The standard theory for gas giant formation is the core accretion model. The detection of a population of short-period terrestrial-mass objects would provide observational evidence for this model (Zhou et al. 2005). The core accretion paradigm is not without its problems however, as it cannot currently explain the large range of orbital eccentricities observed amongst known exoplanets and it requires that gas giants migrate towards their parent stars in order to explain the observed orbital separation distribution.

The detection of Earth-mass planets also critically tests current search techniques – not so much because of the precision of the techniques themselves, which has constantly improved over the last decade, but rather because planet-hosting stars themselves are becoming the fundamental limiting factor. Indeed, because of its very high precision, the Anglo-Australian Planet Search (AAPS; Tinney et al. 2001) has played a significant role in the detection of tiny motions on the surface of Sun-like stars (e.g. Bedding et al. 2007). The study of periodic variations in brightness and surface velocities due to *p*-mode oscillations has been established in the Sun for several decades (e.g. Chaplin & Ballai 2005), and is known as *helioseismology*. These variations constitute a significant source of noise, however: constructive interference between different oscillations leads to peak-to-peak variations as large as 10m/s, which are similar in size to those caused by low-mass planets (albeit on different timescales). We have made significant improvements to our observing strategy used to search for exoplanets by characterising these effects and determining the optimum way to correct for them (O'Toole, Tinney & Jones 2008). Stellar oscillations are just *one* form of stellar variability that affects our ability to detect low-mass planets; stellar activity and convection are others. They each have an effect on different timescales – from minutes and hours to days and even years – and we are attempting to account for each effect as part of an ongoing study.

The AAPS is one of the longest running planet search programs in the world, with over 11 years of observations to date. Prior to this year, we had announced the detection of 29 new extra-solar planets discovered using the so-called "Doppler wobble" technique. The method measures a star's Doppler shift from its spectrum to extremely high precision (velocities at the ~ 1–3 m/s level); using UCLES on the AAT with an iodine cell as a wavelength reference in the case of the AAPS. It has played a major role in the explosion of exoplanet discoveries. Surveys using the technique have discovered the vast majority of the planets known within 200 pc. The AAPS has also achieved arguably the highest long-term stability of ~3 m/s over 11 years, which is required to detect planets further out – i.e. in a similar configuration to our own Solar System – around other Sun-like stars. As discussed above, the precision of AAPS measurements has now reached the level where understanding previously ignored stellar effects becomes important. It also means that the detection of Earth-mass planets is within reach.

Along with a number of low-mass Doppler planet detections (e.g. Udry et al. 2007), the gravitational microlensing detection of the distant 5 Earth-mass exoplanet OGLE-2005-BLG-390Lb (Beaulieu et al. 2006) shows that terrestrial-mass planets do exist. The new frontier in exoplanet research is to find and characterise the local examples. The successful launch of NASA's Kepler mission promises to yield a raft of new low-mass transiting planet candidates by observing one field for around three years. These planets are critical to understanding planet formation and evolution. Doppler searches, however, provide the most accurate estimates of key exoplanet properties: minimum planet mass; orbital period; orbital semi-major axis; and eccentricity. They also differ from photometric transit and microlensing surveys because they specifically target nearby well characterised stars. Hence, they provide robust minimum masses and orbital parameters, and allow critical and independent verification by other groups working in the field. Doppler detections require intensive campaigns on suitably stable host stars; such a programme enables the AAT to produce first-rate science well into the 8-m telescope era.

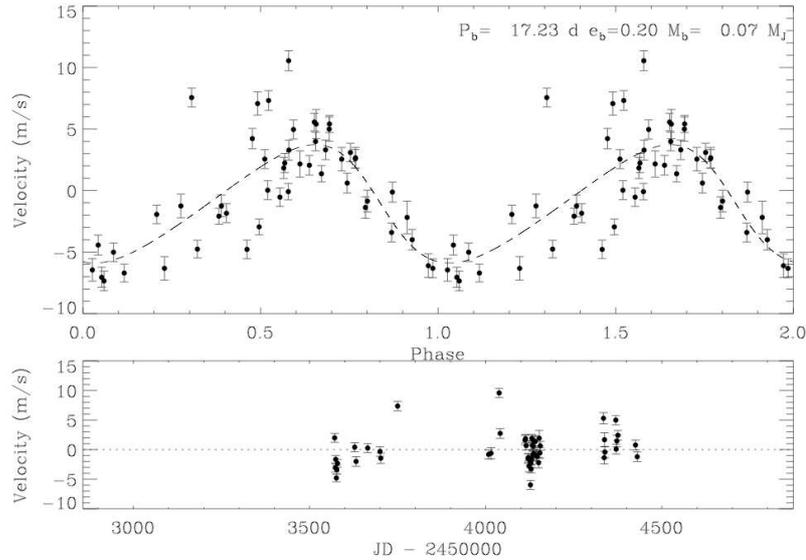

Figure 1

The AAPS team was awarded 48 consecutive nights at the end of semester 2006B to search for low-amplitude terrestrial-mass planets orbiting our most stable targets. The campaign monitored 24 inactive, slowly rotating G & K stars searching for 5–10 Earth-mass planets with periods between 1 and 10 days. (This came after we were granted 64 nights in 2005 over both semesters to search for so-called "Super-Earths" – 10–20 Earth-mass planets in short-period orbits – a campaign that was unfortunately severely hampered by poor weather.)

One of the outstanding results from this "Monster Run" was the discovery of a Neptune-mass (~21 Earth-mass) planet around the G-type star HD16417 (O'Toole et al. 2009a, see Fig. 1). The orbital period of the system is 17.2 days and the orbit appears to be mildly eccentric, although simulations have shown that the eccentricity of low-amplitude signals is difficult to constrain (O'Toole et al. 2009b). The amplitude of the planet's signal is just 4.8 m/s. The planet shows the power of large blocks of contiguous observations to detect the low amplitude variations that are due to low-mass exoplanets. Long observing blocks are vital since, as is the case with asteroseismological measurements, they allow us to improve the window function and integrate up Fourier power. The Monster Run also allowed us to confirm the existence of a 13 Earth-mass planet in a 15.1-day orbit around the star HD4308, an object previously detected using the HARPS spectrograph (Udry et al. 2006). These detections make the AAPS the most successful planet search program worldwide using the Doppler technique, when considering planets detected per star monitored.

In order to produce robust detection constraints from our data, as well as quantify their selection effects, we have Monte-Carlo simulated the data from the Monster Run on a star-by-star basis (O'Toole et al. 2009c). The simulations demonstrate clear differences in the exoplanet detectability from star to star due to differences in sampling, data quality and intrinsic stellar stability. They reinforce the importance of star-by-star simulation when interpreting the data from Doppler planet searches, as originally shown in an earlier study (O'Toole et al. 2009b). The simulations indicate that for some of our target stars we are sensitive to close-orbiting planets as small as a few Earth masses.

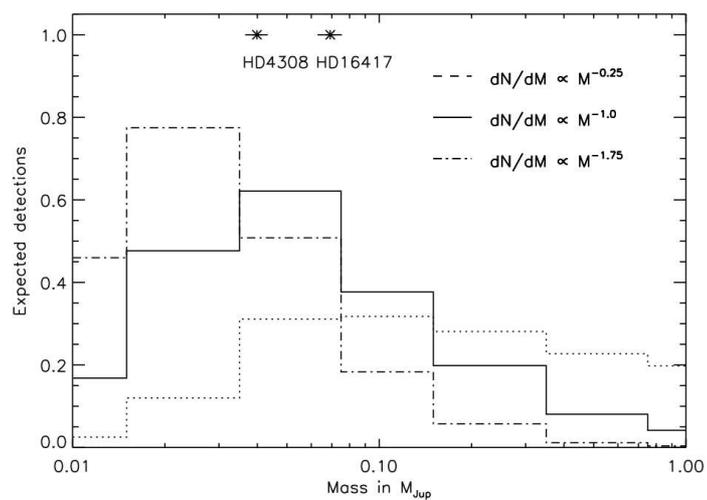

Figure 2

We have used the detectability determined from our simulations to examine how the expected number of detected planets varies when considering different exoplanet mass functions (see Fig. 2). The two low-mass planets present in our 24 star sample indicate that the exoplanet minimum mass function at low masses is likely to be a flat $\alpha < -1$ (for $dN/dM \propto M^{\alpha}$) and that between 15±10% (at $\alpha=-0.3$) and 48±34% (at $\alpha=-1.3$) of stars host planets with orbital periods of less than 16 days and minimum masses greater than 3 Earth masses. Mayor et al. (2009) have claimed that ~30% of stars host such "Super-Earth" planets. While the constraints our data on these 24 stars are able to provide are only modest, they are significant in being the first such estimates that are based on data in which the selection effects have been *robustly* characterised. Our simulations offer a direct methodology to determine an empirical exoplanet mass function. The AAPS has been granted a further 47 contiguous nights over July and August 2009 to bring the total number of targets in the survey to ~60 stars. The results from these observations will improve the constraints presented here.

In the past year we have also announced the detection of a new planet around the M Dwarf GJ832 (Bailey et al. 2009, see Fig. 3). The object is in an almost circular orbit with a period of 9.4 years and mass of 0.64 Jupiter masses. The discovery of this planet – a Jupiter-like planet in a Jupiter-like orbit – demonstrates the importance of our long-term monitoring program of stars similar to the Sun. We are only now beginning to discover planets with periods of 9 or more years, i.e. close to the 11.2-year orbital period of Jupiter. The AAPS has recently been granted large program status on the AAT in order to extend our high precision observations to beyond the orbital period of Jupiter.

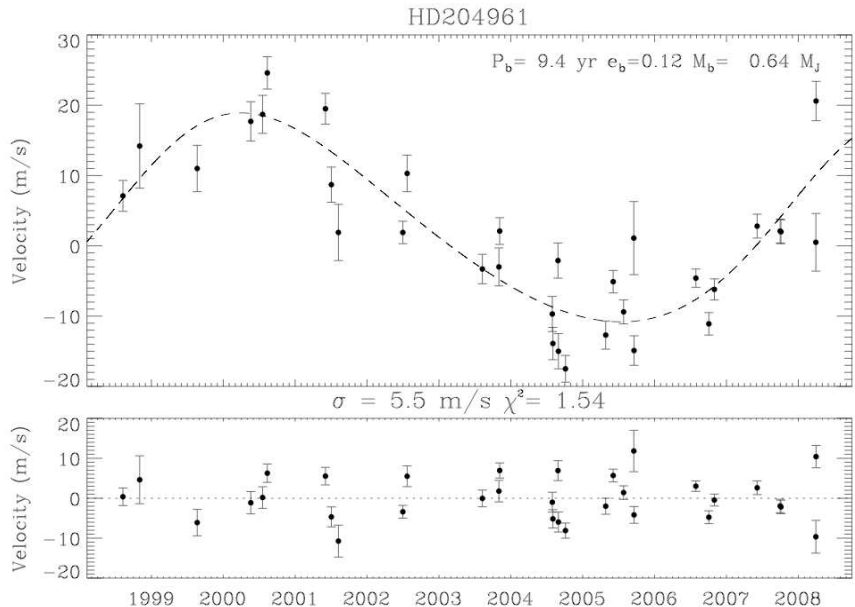

Figure 3

# THE AAOMEGA MINI-SHUFFLE OBSERVING MODE: ACCURATE SKY SUBTRACTION WITH ALL OF AAOMEGA'S FIBRES AT ONCE

Rob Sharp (AAO)

Accurate subtraction of the OH-airglow emission lines imprinted on the red end of astronomical spectra has long proved to be a headache for fibre-based MOS spectrographs. AAOmega, and 2dF before it, are by no means exempt from dealing with the problem. The *nod-and-shuffle* observing technique (Glazebrook & Bland-Hawthorn 2001, PASP, 113, 197) has been implemented with both the 2dF and the AAOmega MOS systems in an effort to provide an observing mode which accurately removes the sky-line signature at the statistical *shot-noise* limit of the data. This technique allows observations over many hours of integration to be stacked, giving very deep spectra (Croom 2004, AAO newsletter, 105, 24). Without the high accuracy subtraction, simple stacking techniques (using the usual dedicated sky-fibre sky subtraction methodology) are typically found to reach a systematic noise floor after several hours of observation, stalling the square-root of exposure time increase in signal-to-noise (or depth sensitivity) one expects when observing in the sky-limited regime.

However, one never gets anything for free. Unfortunately nod-and-shuffle imposes some serious constraints on observing efficiency and so is seldom the correct technique for one's observing program (a clear example of a case where it is wise to discuss your observing strategy with the instrument team before submitting an observing proposal). The single greatest limitation has historically been that one must physically mask-off 50% of the 2dF fibres, preventing their use in MOS spectroscopy and halving the number of targets that can be observed. This masking is required to provide the free space on the CCD into which charge is *shuffled* as the telescope is *nodded* between the A and B positions during an observation.

The excellent image quality of the AAOmega cameras, which deliver the expected fibre PSF profiles with a FWHM of ~3.4 pixels, raises an interesting possibility. The AAOmega fibre pitch (separation between fibres) is 10 CCD pixels, so might it be possible to use all 400 2dF fibres and pack the inter-fibre gaps with the *shuffled* charge? This would place 800 spectra on the CCD at once and remove the need to mask off fibres, with the associated x2 efficiency loss. However an evil spectre raises its head at this point...

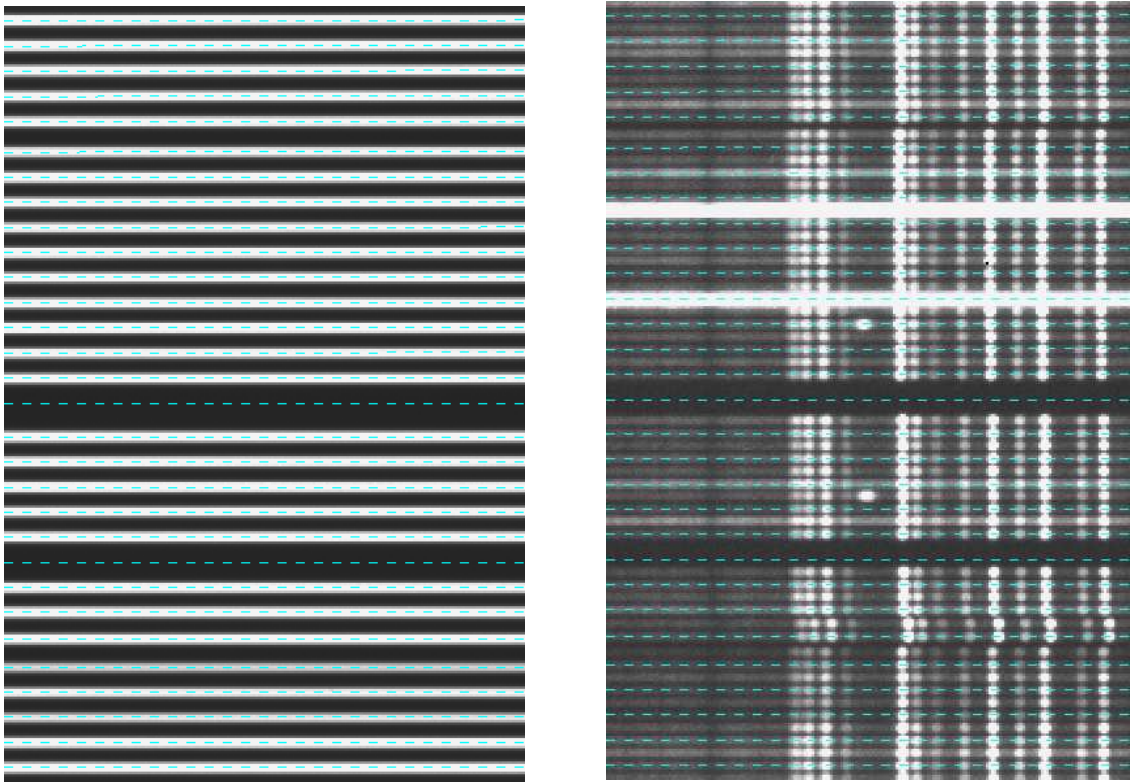

Figure 1: A portion of the raw AAOmega CCD data is shown. On the left we see the *un-shuffled* flat field frame. Fibre centroids are marked by dashed lines. On the right we see the *charge-shuffled* science data. It contains twice as many fibre profiles as the flat field, since both the A and B *nod* positions are present, the B position observations stored in the inter-fibre gaps. Two strong object emission lines can be seen in the region of the CCD shown. These are from the same source, and represent the A and B position spectra from a fibre pair allocated to this target to allow 100% on-source observation.

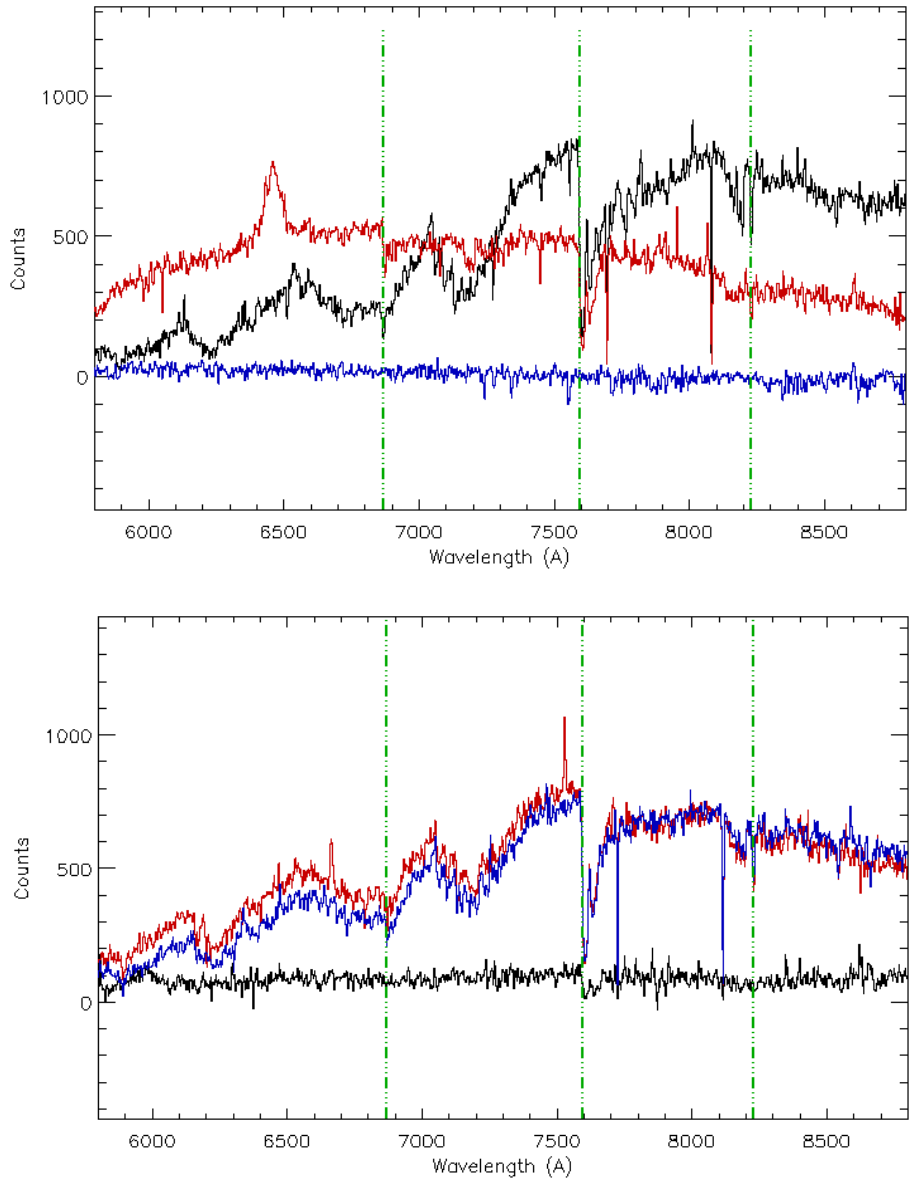

Figure 2: Two sets of fibre triplets are shown. Each triplet is a set of three adjacent fibres on the CCD. Vertical green lines mark the strong telluric Fraunhofer absorption bands (from atmospheric $O_2$). These are multiplicative features that are not corrected for by the additive *nod-and-shuffle* process. In each case there is no trace of the highly structured spectrum of one source as a contamination in the adjacent source, suggesting that fibre-to-fibre cross-talk has been neutralised.

**Fibre-to-fibre Cross talk**

In any system where spectra are tightly packed onto a detector one must always worry about cross contamination of the light from adjacent spectra. The fibre pitch and PSF profiles properties can be easily shown to minimise this problem for AAOmega in MOS mode, provided one maintains a small ($\Delta$-mag<3) spread in input target magnitudes. However, during commissioning of the SPIRAL integral field unit for AAOmega, it was realised that the smaller fibre separation of the SPIRAL system (pitch 4 pixels, with 2.4 pixel FWHM) would require an improved extraction algorithm to account for the fibre-to-fibre profile overlaps. With the implementation of this *optimal extraction* algorithm in the 2dfdr data reduction environment it becomes possible to attempt full 400 fibre AAOmega nod-and-shuffle observations. This mode of observation, in which fibre profiles are *shuffled* into the 10 pixel inter-fibre gaps, is termed mini-shuffling.

The resulting CCD fibre format is illustrated in Fig. 1, while a graphic demonstration of the ability to extract closely packed spectra without cross contamination is shown in Fig. 2. Two sets of spectra are presented; in both cases the spectra are from three adjacent fibres on the CCD. One set shows: **Black** – a low mass star with a strong red spectrum and absorption bands; **Red** – a low redshift quasar with broad smooth continuum and a single emission line (likely Mg$_{II}$); **Blue** – a fibre which was not used in this configuration due to a low target space density (and hence should show no signal). The second trace shows two low mass stars (**Red** and **Blue**), whose spectra bracket a low surface brightness galaxy (**Black**). The two M-star spectra are of the same source, from a pair of fibres allocated to the target to allow 100% of exposure time on source. In all cases, there is no signature of the strong (peak ~800 count) highly-structured spectra cross-contaminating their neighbours.

The data presented here is taken from the service program of Lauroesch et al. (see page 14 of this newsletter). The point source targets were relatively bright for *nod-and-shuffle* observation with AAOmega (r(AB) ~ 19–20); but were observed during bright-of-moon due to scheduling constraints. Since sky gradients from moonlight can limit the accuracy of sky subtraction using dedicated-sky-fibre sky subtraction methods mini-shuffling was used. The sky subtraction accuracy approaches the shot-noise limit as expected for *nod-and-shuffle* observations. A full characterisation of the various AAOmega MOS observing modes is under way. A better understanding of the strengths and limitations of both masked and mini-shuffle nod-and-shuffle observations will not only guide AAOmega users towards the correct strategy for observations, but will also play a key role in the development of the next generation of fibre instruments for ELT facilities in years to come.

**Sky subtraction accuracy**

The ultimate sky subtraction accuracy attainable with mini-shuffling observations can be assessed via a stack of many hours of observations. The observations presented here consist of 3x40 minute exposures and hence 1 hour is spent on target while two hours sky is accrued, one hour in each of the nod positions. For the purposes of demonstration 25 spectra with low source brightness and limited intrinsic spectral structure were selected for summation to produce an effective 50 hour sky observation. Each spectrum is smoothed with a generous kernel and the smoothed spectrum subtracted to remove spectral structure from the source (which were broad line QSO). Figure 3 shows the resulting stack, in which one would expect to see a suppression factor

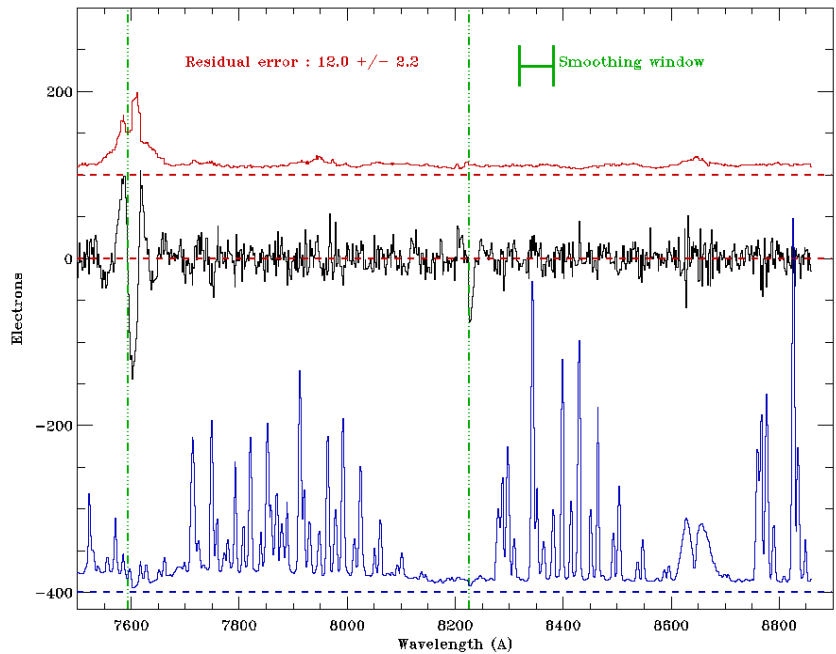

Figure 3: Resulting residual noise after stacking (via a clipped mean) the two hour observations of 25 individual sources. The continuum is first removed via a local smoothing on the indicated scale. The atmospheric Fraunhofer absorption bands are not corrected via nod-and-shuffle and so dominate the residual here. The upper red trace shows the local scatter at each pixel (baseline offset indicated by the dashed line). The lower blue trace shows a fiducial 1% scaling of the sky spectrum. The residual as a percentage of the local sky spectrum is shown in Figure 4.

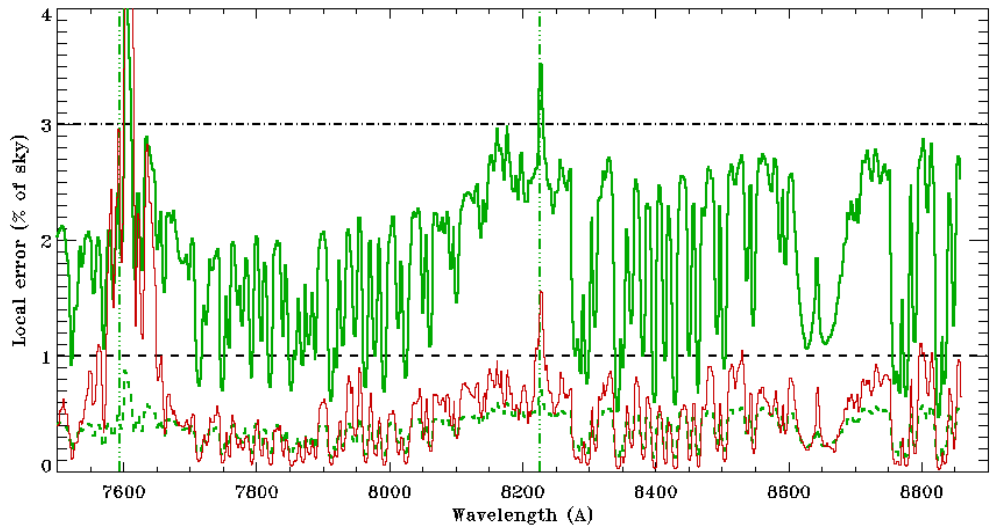

Figure 4: The sky subtraction residual for the 25 hour spectral stack is presented as a percentage of the local sky value. The thin red line shows the local residual error as estimated by the standard deviation of pixel values within the smoothing kernel indicated in Figure 3. The upper green trace shows the shot-noise limits that should be achievable for 1 hour's observation limited by the sky intensity. The lower dashed green trace shows this shot-noise limit scaled for to the total exposure level. The residual levels are consistent with the shot-noise limits within the constraints of the simple statistical tools presented.

of five in the sky residual with respect to the individual component data.

Figure 4 shows the residual sky subtraction error as a percentage of the local sky intensity. The residual error is estimated from the average scatter in pixel values within the same smoothing length given in Figure 3. This simple measure does tend to overestimate the residual level in regions between sky lines while underestimating it for isolated sky features. In regions outside of the atmospheric absorption bands, the subtraction is indeed found to be at the expected shot-noise limit, within the limits of the simple statistical measure used.

**When should I use Nod-and-Shuffle?**

The relative merits of the various observing modes are the subject of ongoing investigations by observatory staff and one should consult the AAOmega support astronomers before deciding on a detailed observing strategy using Nod-and-Shuffle. What follows is our guide to current best practice.

Dedicated sky fibres – the Default option. This mode is correct for most AAOmega projects where optimal signal-to-noise is required (at the expense of the integrity of some sky affected pixels). For target magnitudes r/i~21–20, and fainter for emission line only sources, this will likely generate the best results.

Mini-shuffling – r/i=22–21 continuum sources and long exposures. r/i<20 sources during bright of moon (when moon light gradients might otherwise compromise sky subtraction significantly. Any application where the pixel-to-pixel integrity of data is of paramount importance at the expense of raw signal-to-noise within a spectrum.

Standard Nod-and-Shuffle – Deep, r/i>22/21, multi-night observations of faint continuum sources.

# MEASURING THE QSO-CLUSTER RELATION IN TWO LARGE QUASAR GROUPS WITH AAOMEGA

J. T. Lauroesch, L. G. Haberzettl, G. M. Williger (U. of Louisville, USA), K. A. Harris, R. G. Clowes (U. of Central Lancashire, UK) and the Clowes-Campusano Large Quasar Group Consortium

The QSO phase of galaxy evolution is a relatively brief but highly visible marker of galaxy evolution, which is likely to be environment dependent. QSOs and AGN have significant effects on their host galaxies and surrounding regions on scales of at least hundreds of kpc (Sijacki et al. 2007; Di Matteo et al. 2005), with potentially distinct modes of feedback. Environmental effects are especially important in triggering/quenching star formation through mergers, harassment and gas stripping (Postman et al. 2005). Conversely, the environment around a QSO/AGN host galaxy has a significant effect on its star formation history and the feeding of the central black hole. Indeed, specific models exist linking the QSO luminosity function and the formation rate of spheroids as a function of redshift, allowing specific predictions of galaxy properties (such as galaxy mass functions, star formation rates, etc; Hopkins et al. 2006). The epoch $z\sim1$ is a key point, where the galaxy luminosity function still can be readily probed to faint levels with current telescopes/instrumentation, and environmental effects appear to affect galaxy evolution strongly. While there have been a number of recent surveys which have studied these effects in "typical" regions at $z\sim1$, none have targeted regions with over-densities of bright AGN/QSOs. However it has been known for two decades that there are regions which are over-dense in QSOs (Crampton, Cowley & Hartwick 1989; Clowes & Campusano 1991). Such structures are too large to be virialised, and may reflect large scale primordial density perturbations which have similar enough conditions over a large distance range to produce quasar activity over a short time compared to quasar lifetimes. Since QSOs may signal gas-rich merger environments (Hopkins et al. 2008), and large quasar groups (LQGs) are potentially unique structure markers on scales up to hundreds of Mpc, LQGs could therefore provide a very efficient means to study both quasars and galaxies in a wide variety of environments, from low to high densities. While LQGs are relatively rare objects with a space density at $0.3<z<1.9$ about 3–4 times lower than superclusters at $z<0.1$ (Pilipenko 2007), there are 6 such structures with ~15–20 members found in the 2dF survey. One of these "rich" objects was previously known, the Clowes-Campusano large QSO group field (Clowes & Campusano 1991, 1994), which shows significant excesses in the number of bright QSOs at $z\sim0.8$ and 1.3. With this in mind we have undertaken over the past several years a detailed study of the galaxy populations in the Clowes-Campusano large QSO group field (Haberzettl et al. 2009). This program has included deep UV, optical and near-IR imaging and spectroscopy of a portion of the Clowes-Campusano LQG.

As part of this larger program, we proposed AAOmega observations of ~120 photometrically selected QSO candidates neighbouring (on the sky) large galaxy structures to obtain redshifts to determine the three-space QSO distribution to: (1) confirm that the QSO-galaxy associations are real physical structures, and (2) characterise the QSO populations in this field. The AAT 2dF AAOmega system provides the optimal areal coverage and sensitivity to efficiently obtain the required QSO redshifts. As discussed by Rob Sharp, these observations were obtained using the new mini-shuffling observing technique, and the results provided spectra of even the fainter targeted QSO candidates.

## TCS PROJECT

Steven Lee (for the TCS team – Lew Waller, Keith Shortridge, Anthony Horton, Bob Dean, Minh Vuong, Tim Connors)

Just after morning tea on the 25th November 2008, Bob Dean and I switched off the Interdata Model 70 computer that had faithfully controlled the AAT for the past 35 years for the very last time. There was no great ceremony, but much reverence. This computer was one of the oldest still in active service, its origins in a distant past where a computer was a rare (and very expensive) thing.

First turned on for the AAT project in mid-1972 at the project office in Canberra for use by the engineers and programmers, it was eventually introduced to the telescope in January 1974 where it has remained ever since. It finally left the AAT on March 25th 2009, loaded onto the back of the truck and on its way to the Powerhouse Museum in Sydney where it would be exhibited – and undoubtedly looked on as a curiosity of early computer technology (I can hear cries of "that's a mini-computer! A bit big, isn't it," and "kilobytes – surely that's a mistake. How could you fit any program into so little memory?").

Its specifications were always impressive: a full complement of 64KB of core memory (with 1 microsecond access time), fast 4MHz clock speed and twin 2.5MB (removable pack) disc drives coupled with a potent real time operating system – and it looked like a real computer with a front panel of switches and flashing lights and a paper tape punch and reader. Programmed in assembler and FORTRAN it drove the telescope perfectly, setting the benchmark against which all subsequent telescope control systems would be judged.

Its own success, unfortunately, made it a difficult act to follow. The AAO discussed replacing the computer many times over the years but it was always deferred in favour of another instrument project. After all, a new computer wouldn't add anything to the AAT's capability – it wouldn't make the telescope slew any faster, point any better or track more smoothly – so there was little incentive. The Interdata was also extremely reliable, causing very little time lost over those 35 years – once the initial teething troubles were worked out.

However, it couldn't be put off indefinitely. Bob wouldn't be around for ever to repair it, and it was clear that the technology of 1970 couldn't be pushed into what was needed in today's observatory. Over the years we had performed several upgrades to allow remote access to the telescope. The original interprocessor link (IPL) only worked between the Interdatas in the control room (and was never really used – astronomers in the early years couldn't see the point in having an instrument computer control the telescope), and the upgraded IPL (which used CAMAC as an intermediary between computers) only really allowed our local VAX to communicate with the TCS.

In the mid 1990s, with instrument control shifting away from the ageing VAXs (which had our CAMAC interface), another IPL was needed. This came in the form of a separate computer (one of our standard VME/68xxx/VxWorks crates – known as CCSgate) running a version of Jeremy Bailey's PTCS and talking via ethernet to the outside world and connected to the Interdata via a DR11W DMA interface. This proved quite successful,

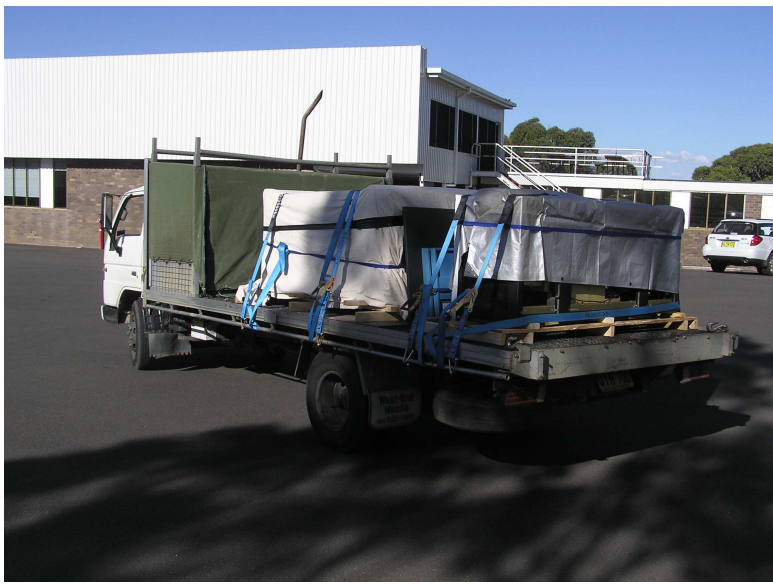

although really pushing the 64Kb memory of the Interdata to the limits.

PTCS – the Portable Telescope Control System – was developed in Hawaii by Jeremy Bailey and others, initially for the JCMT but with the novel intention that it be a truly portable system available for use on other telescopes (Jeremy always had the AAT in mind as a suitable candidate). PTCS would implement all the necessary functions of a virtual telescope, requiring only the hardware layer to be coded for an individual telescope. This has proven to be a success, being used on other telescopes such as UKIRT, the UNSW patrol telescope and now the AAT.

In 2002 we started seriously thinking about a full replacement of the TCS. Jeremy suggested PTCS as the natural choice, an especially attractive option as all our instruments now talked to PTCS, which they thought was the real TCS. This was not an easy choice as there were other options, particularly TCSpk by Patrick Wallace (one of the original programmers for the Interdata TCS). Things progressed slowly (as IRIS2 was taking most of our attention) and it wasn't until September 2004 that a PDR took place and some decisions were made. Unfortunately, this coincided with Jeremy leaving the AAO for the Macquarie Astrobiology department and so Keith was thrown in at the deep end.

The early debates on system design were centred around which computer and operating system to use, and in which language to write the code. The traditional engineering approach is to use a "proper" real-time operating system attached to a well-known interface-friendly CPU. However, some thought that this created a poor user environment, lacking in many of the tools that were desired for creating such a complex system. In the end, a standard Linux installation was chosen running on an Intel CPU, but mounted in a VME crate to allow for sufficient expansion cards. This non-standard approach has paid off, providing both adequate hardware interfacing capability and a rich software environment. Real-time requirements turn out to be no major problem for driving a telescope – the current system runs in "user" mode with no tweaks needed. PTCS was chosen as the core and the rest of the software is written in C++ and C with the GUI in Tcl/Tk, again a contentious choice but one that has worked out well.

The first job for Lew Waller was to understand what it was he had to replace. The original hardware documentation had to be sifted, sorted and correlated until all the digital inputs were traced, all the analogue inputs were characterised and all the outputs defined. It was a testament to Lew's thoroughness that after he'd finished, the telescope hardware interface was better documented than it had ever been.

In parallel with Lew sorting out the hardware, the software group were busy coding up a telescope simulator. Access to the hardware was clearly going to be difficult, and so this step was necessary to ensure that there was little telescope down time needed (and to avoid the programmers having to spend so long at the telescope and away from home!). The simulator has proven to be a valuable resource that is still in use, allowing changes to be made to the software and tested without risk to the working system. It has also shown that the choice of operating environment was a good one as the simulator, and the whole of TCS, runs on just about any Unix-like system, the majority of the TCS coding being done on Linux and Mac laptops.

By the time the hardware was finally brought up to the telescope (in April 2008, long after the hardware was complete as Lew had been taken away to work on another project), the software was happily driving the telescope simulator and most of the necessary features appeared to be working. It was a tribute to both hardware and software efforts that when they were finally connected and the software was instructed to drive the telescope – it did. There were no surprises – it just worked.

On-sky commissioning started in April 2008. The first time the new TCS was asked to drive the telescope to a star was very pleasing as the star landed almost perfectly centred in the camera's field. It then disappeared behind a cloud – something which set the tone for most commissioning nights which followed and delayed acceptance of the new TCS for a long time; it is very difficult to verify a telescope control system when the sky is cloudy. We seemed to get all the cloudy nights; of the 16 nights allocated, we ended up with effectively less than 3 nights where useful on-sky tests could be performed.

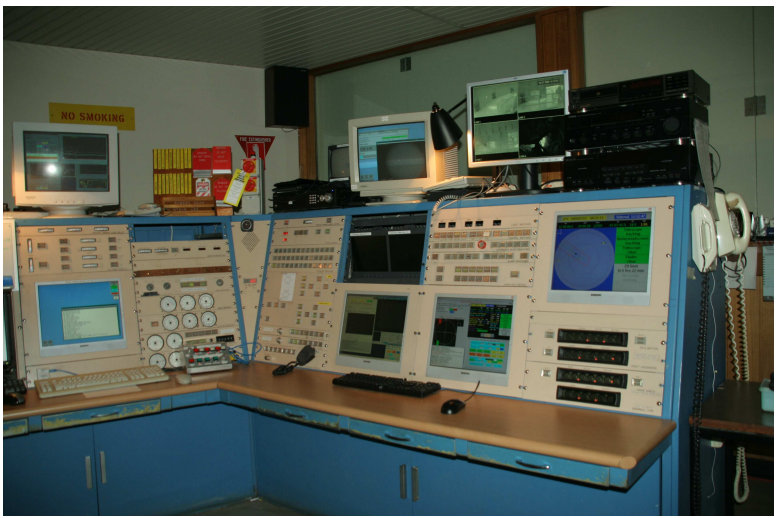

Over the next few months the remaining major deficiencies were sorted out (tricky, given the cloudy commissioning nights) and we started using the new TCS as the default system after the August 2008 commissioning nights (which were mostly cloudy). The new hardware allowed telescope control to be very quickly switched between the old and the new systems – flicking one switch swapped which computer sent control voltages to the motors, while the encoder outputs were sent to both computers in parallel. This allowed comparisons between the 2 systems to be performed very quickly, and provided a backup path should the new TCS be found to be wanting in some necessary function.

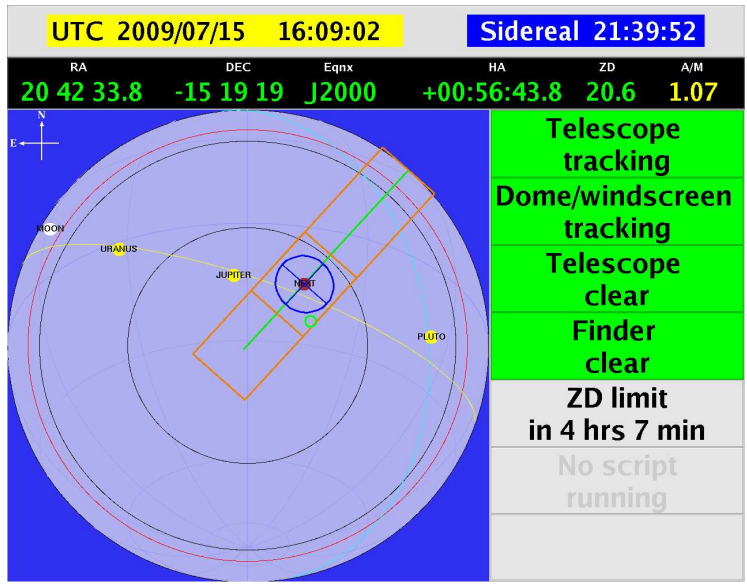

The poor weather we experienced for commissioning meant that the system still had some minor problems which ended up being de-bugged during observing, causing some minor inconvenience but no major problems. The new TCS was rapidly accepted by both night assistants and observers, with the new look and feel appreciated by all. While no new features were added (nor really possible since the old system did everything that a TCS needed to do) the use of a modern computing environment, more powerful computer and graphical user interface meant many things were able to be done in a much more user friendly way; e.g. a graphical display of the position of the telescope, dome and windscreen, along with the sky position of the next observing target, the moon and Milky Way. The new system should also be easier to maintain and, if necessary, extend. In short, we believe that it has been a success and will continue to allow the AAT to be used long into the future. The TCS project is mostly complete now, with everything appearing to work well. There are still some minor features to be finished, but these don't affect observing (and are mainly tidying up and incorporating some of the tools which were used for testing and are worthwhile keeping). After the start of the TCS project the AAO received money from the government for infrastructure upgrades, which included replacing the old encoders and telescope drive motors. So far several new encoders have been installed on the telescope, with the remainder to follow shortly; while new drive motors and amplifiers are still being tested. These changes were not a part of the original TCS project, but having the new system will make it much easier to incorporate them.

The final stage of the TCS project is to remove the Interdata and all the old hardware. During the November 2008 commissioning run we made the irreversible decision to begin removing the old system. The Interdata is now gone, but as yet the old Rate Generator (the major hardware interface between the Interdata and the telescope) still remains in the control room. At the time of writing this, the controls from the console have mostly been removed and the new terminals installed in their place, making it look a very different console (see picture on page 16). The process of removing the final wiring is slow as we need to make sure that no unexpected problems occur (we've already encountered one surprise – the removal of the old timing system caused the tube extension encoder to stop working as there was an undocumented timing signal being used).

Finally, the area once occupied by the Interdata and Rate Generator hardware will leave a space in the control room. The observatory has yet to decide what to do with this area, but I'm asking for an "overflow" tea room. Observers are asked to comment on this when they fill in their observer's report form after their run.

# THE INTERDATA MODEL 70 COMPUTERS OF THE AAO
Graham Bothwell

Visiting the AAO in October 2008 and being shown around by Robert Dean brought back many fond memories. My sixteen years at the AAO, including years of planning, construction, commissioning, and operations, were profoundly formative years, providing a wealth of experience that would have been difficult to acquire elsewhere.

The highlight of my recent few hours at the AAO was to discover that the Interdata Model 70 computer that controlled the telescope for over 30 years was about to be retired. Thus it seems appropriate to reminisce about the origin of these machines, why they were chosen, and some of the early experiences. Working on the AAT in those early days was special in that the AAT was one of the first telescopes to have a computer. The only telescope we knew at the time that had computer control was the 84-inch on Mauna Kea.

So what did the AAT planning team expect from a computer? The late 1960s and early 1970s were an era of mainframes, with a range of smaller machines for industrial control applications. Because the concept of a computer controlling a telescope was novel, we let a design contract with what was then GEC, in Leicester, and I spent the whole of 1970 based at their factory while Geoff Ohlsen and his team worked on developing a hardware and software design. There was no concept of the massive software applications that even small computers of today can support. We anticipated that software would be written in assembly language, probably without an operating system. Reliability was a big issue, so for mass storage the design was based on a magnetic drum with a capacity of around 200Kwords.

However, as we prepared to go to tender, a couple of things emerged that resulted in a very different kind of computer system than had been envisaged even a few months earlier. Most significantly, the first of what became known as "minicomputers" appeared. And there were multiple vendors, including Digital Equipment Corporation (DEC), Data General, Varian, Hewlett-Packard, and Interdata.

In our assessment of the alternatives, it was clear that Interdata offered several features different from any of the other contenders. One was that Interdata proposed the just-released Diablo interchangeable disc drive, with the then-seemingly huge capacity of 2.5 Megabytes. Would it be sufficiently reliable? That aspect was argued at great length, but the vast difference in capacity between this and the non-interchangeable magnetic drum was clear, and the newer device seemed to be the way the industry was heading.

Another factor was software: Interdata had both a Fortran compiler and a multitasking real-time operating system (RTOS). That is, we didn't need to submit ourselves to the rigours and limitations of assembly language, but could do much of the coding in a higher-level language and have the benefits of a sophisticated operating system. Looking back just a few years later, we were incredulous that other vendors did not at the time have these features.

A less-obvious but highly elegant attribute of Interdata was the industrial-style interface using ferrite cores for electrical isolation of digital input signals, such as for telescope encoder readings. (Later, during the commissioning period on the mountain, I discovered how baffling it can be when these interfaces are accidentally connected the wrong way round – the behaviour is far from obvious.)

These factors put the Interdata in a class on its own, although they also gave it the appearance of greater risk. But overall Interdata simply offered more for the investment than did others. And so it was.

It was as if we had purchased a small version of a mainframe! In a sense that was true. The Interdata architecture, including the instruction set, was modelled after the then-common IBM 360-series mainframe. And the Interdata gave us an unusually large memory capacity, 64 Kilobytes!

The Interdata had an attribute that was always something of an oddity – punched cards. We required that all software be available as source code, and it came as punched cards, including the Assembler, the Fortran compiler, and operating system (the latter occupied 10 drawers, 20,000 cards). In my responsibility for the system aspects of the computer, the availability of source code provided numerous opportunities for customisation, such as greatly speeding up compilation and assembly by reading and writing multiple blocks of data instead of one block at a time; special assembler and compiler commands; numerous special calls to the operating system; and real-time interface drivers. And it was most unusual in that era for the AAT software engineers to be using punched cards rather than paper tape for software development.

Originally, Interdata was represented in Australia by a company called Datronics, and they provided substantial special engineering assistance for the AAT. Examples include a high-speed video display interface, containing over 100 integrated circuits, designed and built by Peter Horn, and numerous operating system enhancements by Bruce Williams. The company was keen to include local content, so the cabinets were made in Australia, at least for that initial system.

While the telescope control computer was one of a kind, we developed a more-or-less standard instrumentation computer configuration, of which several were built, including those installed at Epping. An instrumentation system could be assembled for approximately $100,000. Use of the Camac hardware interface for astronomical instrumentation was another innovation for the AAT, and one that provided considerable flexibility. But that is another story.

Interdata, whose manufacturing facility was in New Jersey, was eventually absorbed by Perkin-Elmer, and their computers were in due course overtaken by other architectures. While Interdata/Perkin-Elmer computers ranged up to large 32-bit virtual-memory machines, only the Interdata Model 70 was used at the AAT.

Even during the early years of the AAT, our use of Interdata was regarded as unusual. While Interdatas were used in British astronomy, originally for compatibility with the AAT, these machines were never used as extensively by industry as some other machines, such as DEC's PDP-11 series, which became so ubiquitous in the 1970s and 1980s. However, at the AAT the Interdatas were highly successful, and showed that they did indeed provide what they originally promised. And having one of them still operational to drive the telescope after so many years is a tribute to the highly reliable design, as well as to those members of staff at the AAT, such as Robert Dean, who maintained them.

---

*Biographical note:* Graham Bothwell was an AAO staff member between 1969 and 1985. He then worked for NASA's Jet Propulsion Laboratory until retiring from JPL in October 2008. He lives with his wife Mary in Pasadena, California.

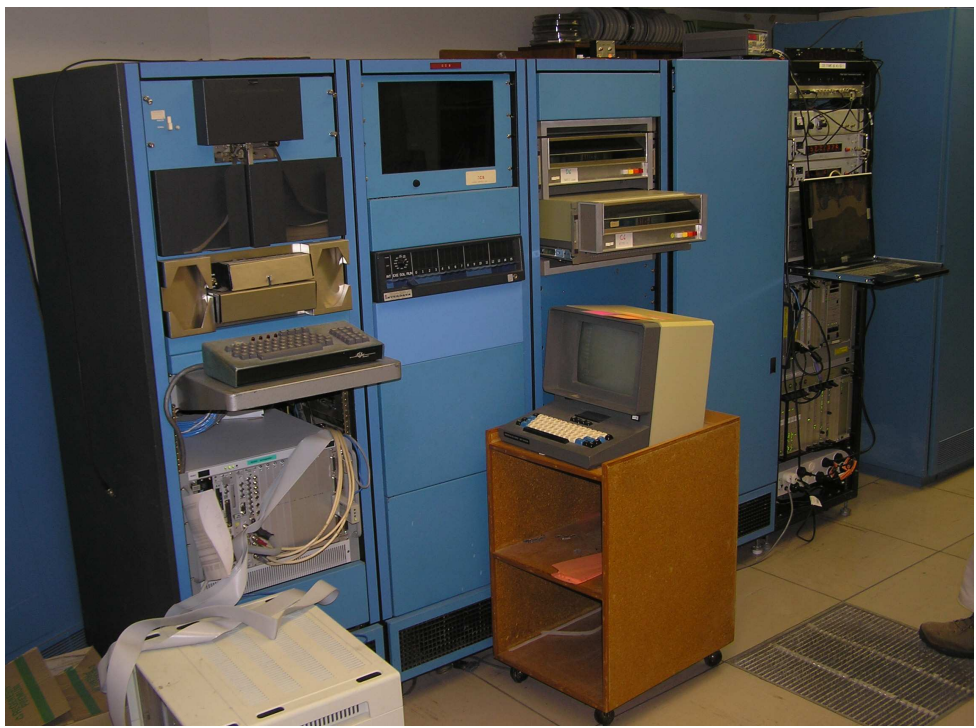

AAT telescope control computer in 1973, in the midst of software development, installed temporarily in a shopfront in suburban Canberra, prior to installation at the telescope site.

# THE INTERDATA MODEL 70 – COMPUTER CONTROL SYSTEM – CCS

Robert Dean

The Interdata Model 70 computer was manufactured by Interdata Computer Systems at Oceanport, New Jersey. The company was originally incorporated in September of 1966 with the object of becoming the leader in the then fast growing and competitive minicomputer market. Out of the various models manufactured by Interdata, the Model 70 was chosen as the system to control the AAT. Made at the beginning of the 1970s, it had very impressive specifications for its time:–

Memory: 64kbytes of Directly Addressable Magnetic Core Memory of 3 wire 3D design with 1.0 microsecond cycle time and 300 nanosecond access time. A notable attribute of core memory was that memory content was not lost with the removal of power.

Disk System: Comprising 2 Diablo disk drives using 2.5 Mbyte removable disc packs, a total of 5Mbytes of storage were available.

Processor: Made from discrete DTL/TTL devices with a Master Clock Generator running at 16 MHz producing a System Clock speed of 4MHz, the Processor utilized a Micro Control Store which stored a micro program (in Read-Only-Memory – ROM) to execute the users' instructions. It had 113 native instructions including fixed point and floating point. Instruction execution times varied from 1 microsecond for Register type instructions to 54 microseconds for floating point multiply instructions. There were 16 hardware general registers available.

Selector Channel: Typically called "the SELCH" it was a standard Direct Memory Access device allowing data transfers up to 2000Kbytes/sec for interfacing to discs, magnetic tape drives, VDUs etc.

Digital Multiplexor: This device was quite unique and allowed the processor to control and monitor digital lines. Using a single controller, the system could monitor and control up to 2048 I/O lines. It used a biased core technique (like core memory) to achieve this. This ensured absolute DC isolation from the outside world giving excellent common mode transient response and DC offset capability.

Operating System: The Real Time Operating System (RTOS) was a true real time system. It maximised the machine utilisation by providing a multi-programming capability for interleaving the execution of programs and overlapping I/O operations. Flexibility was enhanced by device independent programming through logical device assignments and automatic acknowledgement of interrupts. AAO staff provided much of the Operating System support and wrote many of the systems device drivers.

Programming Languages: The Interdata computers came with a Fortran IV and Assembler Language compilers. There was even a Basic Interpretor.

Other Devices : The CCS had a large number of external hardware interfaces connected to it - many of which were designed and built by the AAO. These included autoguider interfaces, VDU interfaces, line printer and serial terminal interfaces, plotters, card readers, paper tape punch and reader, Time-of-Day clocks, Memory Protect controllers, external interrupt controllers, interfaces for CAMAC and instrument interfaces e.g. IPCS, Coude Rotator. A very special interface was developed for the CCS to enable it to be accessed via ethernet by other computer platforms. This utilised a VME system with special software written for both the Interdata and VME systems. It was most likely the only Interdata Model 70 computer in the world to have ethernet capability.

The Interdata Model 70 computer systems were the mainstay of the computing systems for the AAO for the first ten years of operation. Besides the CCS system for the controlling of the AAT, two more systems were in use at the AAT for instrument control, data taking and data reduction. These were generally referred to as ICS1 and ICS2. Gradually, other computer platforms were introduced – the instrumentation computers remaining operational for about twenty years and the CCS surviving for thirty seven years. Over the years from 1972, the Interdata computers have been maintained internally by AAO staff and have proven to be the most reliable systems the AAO has had, both in terms of hardware dependability and operating system robustness.

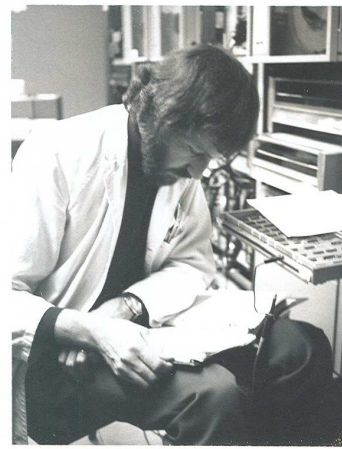

.

## HOW WE OVERCAME GREAT BARRIERS (MOSTLY)
Fred Watson

If you're going to hold an international workshop in a clichéd (some might say iconic) location, you might as well give it a clichéd title. Thus was a recent workshop at Palm Cove on the Great Barrier Reef graced with the name 'Overcoming Great Barriers in Galactic Archaeology'. Groan. But the meeting itself threw clichés out of the window, and got on with some serious business in progressing our understanding of the history of the Galaxy and the Local Group – evocatively known as 'galactic archaeology'.

Organised by the Anglo-Australian Observatory in association with the Astrophysical Institute Potsdam, the Australian National University, and the University of Sydney, the workshop aimed to provide a forum in which our current knowledge could be assessed and future strategies defined. It had the pleasant additional function of celebrating the contributions of two of Australia's most eminent practitioners in the field, Mike Bessell and John Norris of the ANU. As befitted the tranquillity of Palm Cove, the format was relaxed – which was just as well considering it rained every day. However, the workshop unashamedly reverted to cliché on its final day with a hugely popular reef cruise to sunny Michaelmas Cay.

For the Workshop's 40 attendees, there was much to discuss. No-one had any doubts as to the timeliness of the meeting, with large-scale stellar spectroscopic surveys such as the Sloan Extension for Galactic Understanding and Exploration (SEGUE) and the RAdial Velocity Experiment (RAVE) on the UKST now yielding million-star datasets. Moreover, new stellar surveys are on the horizon using instruments such as the ANU's SkyMapper telescope, the LAMOST telescope in China, and the HERMES instrument being built for the Anglo-Australian Telescope. Space missions such as Europe's GAIA and Japan's JASMINE will also yield massive new datasets within the next decade. We clearly stand at a significant moment in the development of galactic astronomy, rivalling the recent leap in precision cosmology that has resulted from large-scale galaxy redshift surveys and the exploration of the cosmic microwave background radiation.

So where are the great barriers that need to be overcome? Principally, they lie at the interface between our best theoretical understanding of the formation and evolution of galaxies, and what we observe. While it is true that on the broadest scale, the standard cold dark-matter paradigm will produce galaxies a lot like our own, the devil is in the detail. And in contrast with the situation only a decade ago, when radial velocities and physical parameters were known for only a few thousand stars in our Galaxy and a few dozen beyond, we are now data-rich – and getting richer – in this field. Thus the challenge has been passed back to the theorists, who are now having to build galaxies using much more sophisticated chemo-hydrodynamical models.

In recent years, our understanding of the overall structure of our own Galaxy has gone through a revolution. The traditional components of disc, bulge and halo have fragmented into thin and thick disc populations, with a radial metallicity gradient in the latter (i.e., a systematic change in the proportion of elements other than hydrogen and helium), and dynamically separate inner and outer halo components.

Details of the bulge and its origin remain mysterious, and are only now being probed with new surveys such as the Bulge RAdial Velocity Assay (BRAVA) on the CTIO 4-m telescope. This project has yielded strong confirmation of the presence of a bar in the centre of our Galaxy, orientated at 20 deg to our line of sight, and suggests that the bulge's kinematic uniformity results from a single population of stars.

From the overall properties of our Galaxy, the meeting turned its attention to the individual stars and star clusters contained within it, and here the great barriers seemed to get significantly more reef-like. Questions ranged from whether the Sun is, in fact, a solar-type star (in terms of our understanding of its chemical composition) to the failure of big bang nucleosynthesis calculations to reproduce the observed abundance of Lithium 7 in ancient halo dwarf stars. Few clues seem to be available to solve this 'lithium problem', but they are currently being explored in detail.

Chemical clues that disintegrating ultra-faint dwarf spheroidal galaxies may have fed stars into the outer halo of our Galaxy were also explored, and found plausible. These virtually invisible dwarf galaxies have been discovered by clever analysis techniques around both our own Galaxy and M31, and their metallicity signatures closely match those found in the outer halo. Some of their stars are very metal-poor indeed (i.e. unpolluted by heavier elements).

The idea of 'chemical tagging' to reveal common origins in disparate groups of stars in our Galaxy also found favour in discussions of stellar streams in the solar neighbourhood. For more than 40 years, we have recognised that streams of stars passing through our vicinity can be identified by their common velocity, but with the new data sets, much more structure in the local velocity field is revealed. How many of these streams are the debris of tidally-disrupted globular clusters and dwarf spheroidal galaxies? Or are many of them simply due to resonance effects in the rotation of the galactic bar and disc?

To help in disentangling such problems, the detailed measurement of the physical properties of very large numbers of stars is essential. Metallicity, surface gravity, effective temperature and so on, are required for us to understand the origin of individual stars, and will allow us to build up a population census of the Sun's neighbourhood that is entirely unprecedented.

To do this requires sophisticated instrumentation, and the workshop ended with the promise of exactly that being provided – in particular by HERMES. The lively discussion as to exactly what elements should be probed by this exciting new instrument inspired confidence in the community's continuing zest for a better understanding of every nook and cranny of our Galaxy.

As Tim Beers (MSU) and Daniela Carollo (ANU) wrote in their wrap-up of the workshop (*Nature Physics*, **5**, 463, 2009), 'May the great barriers to Galactic Archaeology continue to fall, and the magnificent Great Barrier Reef, from which the conference drew its name, continue to live on.' Clichéd or not, I think we can all drink to that.

*Overcoming Great Barriers in Galactic Archaeology – an AAO Workshop and Celebration of the Contributions of Mike Bessell and John Norris*; Palm Cove, Qld, 5–8 May 2009; http://www.aao.gov.au/conf/palmcove/

Presentations can be downloaded at: http://www.aao.gov.au/conf/palmcove/Participants.htm (click on participants' names)

Scientific Organising Committee:

Matthias Steinmetz (AIP) (Chair), Joss Bland-Hawthorn (Sydney), Gary Da Costa (ANU), Annette Ferguson (Edinburgh), Ken Freeman (ANU), Connie Rockosi (UC Santa Cruz), Kim Venn (Victoria), Fred Watson (AAO).

Local Organising Committee:

Fred Watson (AAO) (Chair), Stephen Marsden (AAO), Marnie Ogg (Thrive Australia) (Consultant), Helen Woods (AAO).

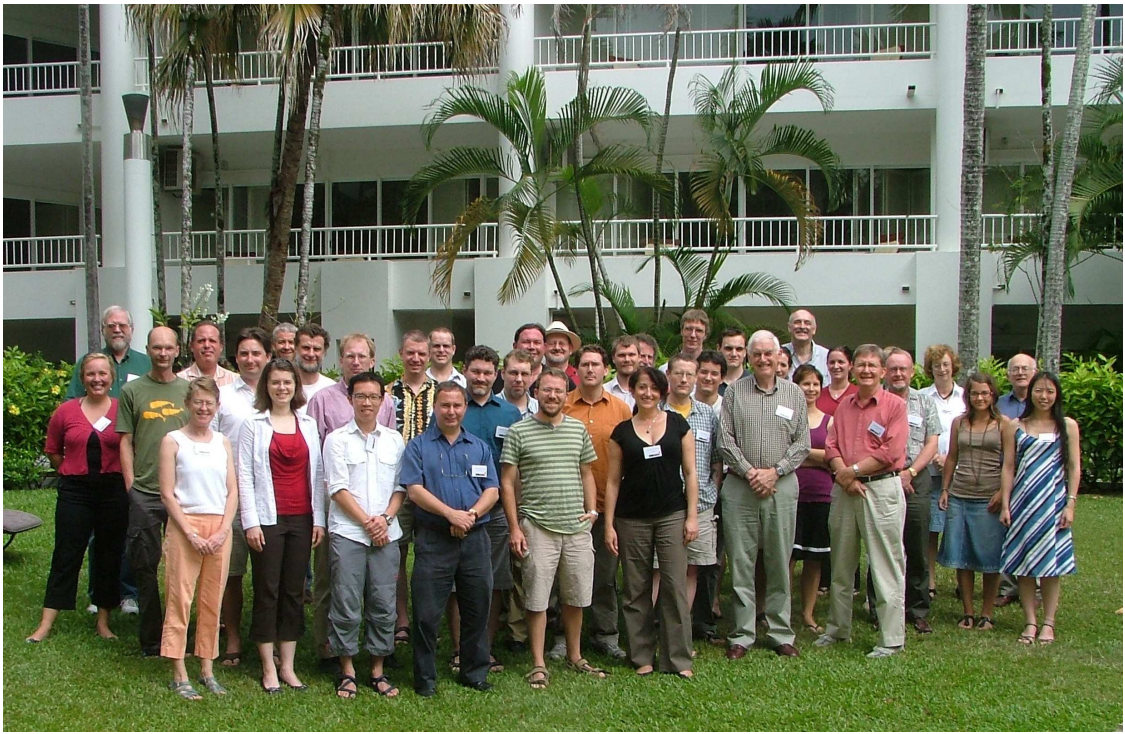

Trying hard not to look as though they are on holiday, attendees at the 'Great Barriers' workshop at the Angsana Resort, Palm Cove, contemplate the future of galactic archaeology with due stoicism.

## AUSGO CORNER
Stuart Ryder (Australian Gemini Office, AAO)

**Semester 2009B**

In this semester we received a total of 35 Gemini proposals, of which 22 were for time on Gemini North or exchange time on Keck or Subaru, and 13 were for time on Gemini South. This was a ~50% increase on recent semesters, and coupled with a 40% reduction in the number of hours available to ATAC in 2009B in order to redress a growing imbalance in partner share of telescope time used, has resulted in the highest-ever oversubscription factors for ATAC time on Gemini: 4.53 for Gemini North (including Keck and Subaru exchange time), and 3.18 for Gemini South. At the ITAC meeting Australia was able to schedule one classical run each on Keck and Gemini South, with 11 more programs going into Bands 1–3.

For Magellan we received 9 proposals maintaining a healthy oversubscription of 2.93. Semester 2009B marks the beginning of the 4 semester extension to Magellan access enabled by Astronomy NCRIS Strategic Options Committee (ANSOC) funding. The consistently high demand for Magellan time, together with the surge in Gemini proposals in a semester when no new capabilities were being offered appears to validate ATAC's recent decision to have the Magellan proposal deadline one week later than Gemini.

**GMOS Imaging Contest**

As its contribution to the International Year of Astronomy, AusGO has been running a contest for Australian high schools to win one hour of time on Gemini South to image their favourite target in multiple filters with GMOS. Although we received just six entries from four different schools by the 1 May application deadline, all were of excellent quality, and the judging panel comprising professional astronomers, science educators, and journalists faced a difficult task in selecting just one winner and two runners-up. After considering both the scientific case presented and aesthetic appeal the judges announced the winning entry to be that of Daniel Tran, a Year 10 student at PAL College in Cabramatta, NSW, who proposed to observe the planetary nebula NGC 6751, nicknamed the "Glowing Eye". The runners-up were Forest Lake College in Queensland, and Sacred Heart College in Victoria, who suggested Eta Carinae and the planetary nebula SMC N70, respectively.

The Phase 2 program to observe Daniel's target has now been prepared and sits in the Band 1 queue, so observations are likely to take place in the next couple of months. In the meantime all three schools will get to host a "Live from Gemini" interactive event from the Gemini control room, where a Gemini staff member will introduce them to the Gemini telescopes and observing process. Once the winning target has been observed, the data will be processed by AusGO staff and the multiple filters combined to yield a spectacular full-colour version to be unveiled at PAL College later in the year. To follow the action of this IYA event please visit http://ausgo.aao.gov.au/IYAcontest/news.html for regular updates.

**AGUSS**

Each year since 2006 AusGO has offered talented undergraduate students enrolled at an Australian university the opportunity to spend 10 weeks over summer working at the Gemini South observatory in Chile on a research project with Gemini staff. At its April 2009 meeting, the AAL Board considered a report from the Australian Gemini Steering Committee on the outcomes of the Australian Gemini Undergraduate Summer Student (AGUSS) program so far, and endorsed the continuation of its sponsorship of this scheme. Applications for the 2009/10 program close on 31 August 2009, so if you know of any students in Australia who might be interested then please draw their attention to http://ausgo.aao.gov.au/aguss.html.

**Magellan Fellowships**

As part of the agreement with the Carnegie Observatories for access to the Magellan telescopes, the AAO employs two Magellan Fellows who are seconded to Las Campanas Observatory for two years providing observer support and carrying out their own research, followed by a third year of full-time research at an Australian host institution of their choice. The first two Magellan Fellows, David Floyd and Ricardo Covarrubias, are about to complete their time in Chile; David will spend his final year at the University of Melbourne, while Ricardo will be working at the AAO. Two new Magellan Fellows have recently been appointed.

Francesco Di Mille obtained his PhD in astronomy at the University of Padua in Italy in 2007, and is currently a resident astronomer at the Asiago Observatory in Italy. His research interests include active galactic nuclei, and studying the physical conditions and motions of gas around the AGN using optical integral field spectroscopy. He is also involved in searches for extragalactic novae. In his final year Francesco will be based at the University of Sydney working closely with Prof. Joss Bland-Hawthorn.

Shane Walsh obtained his PhD from the Research School of Astronomy and Astrophysics at ANU in 2008.

During his PhD, Shane discovered a new dwarf satellite galaxy of the Milky Way called Bootes II by sifting through data from the Sloan Digital Sky Survey. He plans to use Magellan to follow up new candidate dwarf galaxies which emerge from the SkyMapper surveys at Siding Spring. In his final year Shane will be hosted by the new Curtin Institute of Radio Astronomy at Curtin University in Perth working with Prof. Steven Tingay.

Congratulations to Shane and Francesco!

**Gemini/Subaru Science Meeting**

The Subaru and Gemini Observatories held a Joint Subaru/Gemini Science Conference in Kyoto, Japan from 18–21 May 2009, which was followed by a Gemini Users meeting on 22 May. Travel grants from AusGO helped ensure that Australian users of both Gemini and Subaru (via the time exchange program) were well-represented at these meetings with Scott Croom (U. Sydney), Karl Glazebrook, Lee Spitler, Sarah Brough (Swinburne), Stefan Keller, Chiaki Kobayashi (RSAA) and Stuart Ryder (AusGO) all in attendance. Despite the disappointment of the announcement on the opening day by the Gemini Director Doug Simons that the Gemini Board had been forced to terminate the WFMOS project due to funding limitations, it became clear throughout the meeting that the Gemini and Subaru user communities have a lot more in common than a shared desire to see WFMOS proceed on the Subaru Telescope. Indeed, the momentum built up by negotiations over the significant amount of time swaps between the two communities which putting WFMOS on Subaru would have necessitated makes it likely that the keenly-contested 5–6 nights of exchange time currently offered on each telescope will be expanded in the near future.

**Henry Lee Visit**

Whenever a Gemini or Magellan Observatory staff member comes to Australia, AusGO staff are always keen to learn about the latest developments from them, and where possible to organise workshops for the Australian community. A recent visit by Dr Henry Lee, a Gemini Science Fellow at Gemini South to attend the AAO/ATNF "Galaxy Metabolism" workshop provided just such an opportunity. Henry was invited to give a presentation at the AAO on 29 June 2009 entitled "Demystifying the Gemini Queue", in which he outlined just how the Gemini queue works; using Observing Condition constraints to one's advantage; and frequently-made mistakes in Phase 2 preparation. Astronomers at Swinburne University and at RSAA were able to join in by video-conference (see photo below), and in all about 20 participants got to learn more about the Gemini queue system and quiz Henry about it. Henry's presentation can also be downloaded from http://ausgo.aao.gov.au/events.html. We are grateful to Henry for giving up his time to help us all get even more out of Gemini.

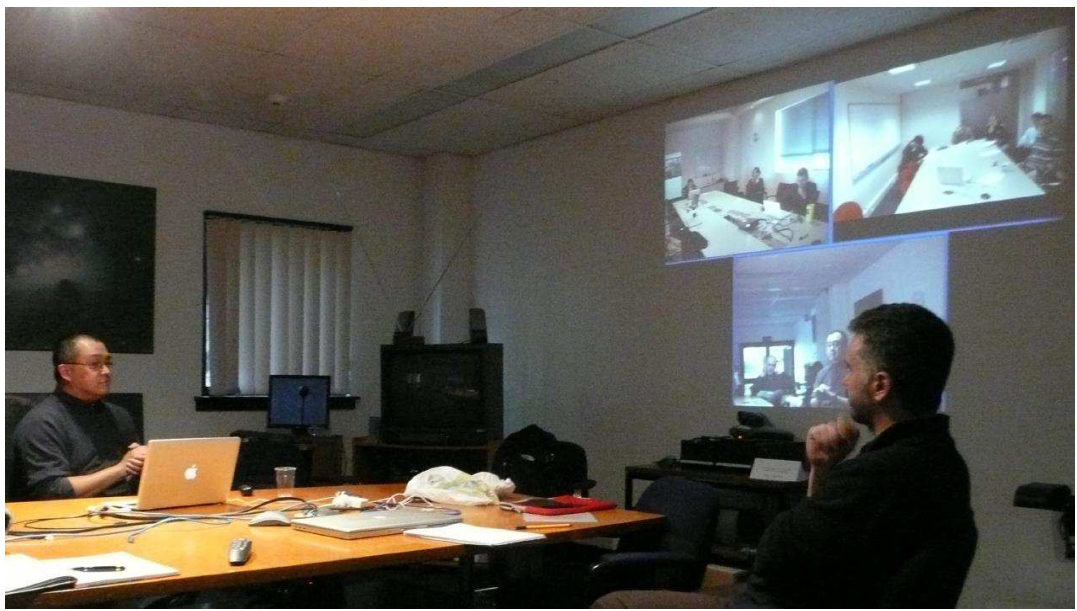

Gemini South astronomer Dr Henry Lee (left) reveals the inner workings of the Gemini queue system to AAO staff including AusGO Deputy Gemini Scientist Dr Simon O'Toole (right), as well as participants at RSAA (top left image on screen) and at Swinburne (top right).

**A SLICE OF THE UNIVERSE: THE FINAL REDSHIFT RELEASE OF THE 6DF GALAXY SURVEY**

Heath Jones (AAO), Matthew Colless (AAO), Mike Read (ROE), and the 6dFGS Final Redshift Release Team[1]

It might have looked like an April Fool joke but in fact it wasn't.

On April 1 this year the 6dF Galaxy Survey (6dFGS) made its Final Redshift Release (DR3) of spectra, images, and redshifts. The DR3 release supersedes all earlier incremental releases and completes the collection and processing of the main data products from 6dFGS. Various additional value-added 6dFGS data products will be the subject of future releases, and ongoing work with 6dFGS now centres on its peculiar velocity survey of 11,000 galaxies.

The 6dF Galaxy Survey is a spectroscopic survey of the entire southern sky with $|b|>10°$, made with the 6dF multi-fibre spectrograph on the Anglo-Australian Observatory's 1.2-metre UK Schmidt Telescope. DR3 redshifts and spectra are accessible through the 6dFGS Online Database at http://www-wfau.roe.ac.uk/6dFGS . The new 6dFGS web site (http://www.aao.gov.au/6dFGS) is the main source of the latest survey news, and contains links to the database, publications, and presentations (Figure 1). There is also a link to the late Prof Tony Fairall's 6dFGS Atlas and a collection of 3D images and fly-through movies.

---

[1] The 6dFGS Final Redshift Release Team:

Heath Jones, Mike Read, Will Saunders, Matthew Colless, Tom Jarrett, Quentin Parker, Anthony Fairall, Thomas Mauch, Elaine Sadler, Fred Watson, Donna Burton, Lachlan Campbell, Paul Cass, Scott Croom, John Dawe, Kristin Fiegert, Leela Frankcombe, Malcolm Hartley, John Huchra, Dionne James, Emma Kirby, Ofer Lahav, John Lucey, Gary Mamon, Lesa Moore, Bruce Peterson, Sayuri Prior, Dominique Proust, Ken Russell, Vicky Safouris, Ken Wakamatsu, Eduard Westra, Mary Williams

The 136,304 spectra of the 6dF Galaxy Survey have yielded 110,256 new extragalactic redshifts and an overall catalogue of 125,071 galaxy redshifts. A paper accompanying the DR3 release (Jones et al. 2009) documents the final status of the survey, and has been summarised in an earlier AAO Newsletter (no. 112, Aug 2007). The original survey paper (Jones et al, 2004) provides more detailed descriptions of sample selection, observational techniques and the survey generally. All of these papers (and many others) are available at http://www.aao.gov.au/6dFGS/Publications .

The 6dF Galaxy Survey final redshift release concludes the first chapter in the 6dF story since the instrument was built a decade ago. The success of the Galaxy Survey owes much to the combined efforts of the instrument builders, commissioning scientists, observers, redshifters, and all who made this ambitious survey possible.

**References:**

Jones et al, 2009, MNRAS in press (http://arxiv.org/abs/0903.5451)
Jones et al, 2004, MNRAS 355, 747

Figure 1: Screenshot of the 6dF Galaxy Survey website.

## SUMMER STUDENTS

The AAO runs a twice yearly fellowship programme to enable undergraduate students to gain 10–12 weeks of first hand experience of astronomical related research. The current crop of students are from the northern hemisphere.

Caroline Yeomans arrived in June, having completed her second year of Physics undergraduate masters course at the University of Oxford. Working under the supervision of Simon O'Toole, Caroline is analysing the ensemble properties of hot subdwarf stars in Data Release 7 of the Sloan Digital Sky Survey, in particular the effective temperature, the surface gravity and the helium-hydrogen ratio. She has already completed the first stage which involved identifying SdB stars in a colour-selected sample of candidates with spectroscopic data.

Marc Etherington has just finished the second year of an undergraduate masters degree in Physics at the University of Durham. He is working under the supervision of Jon Lawrence on a project involving the characterisation of photonic crystal fibres and femtosecond laser-written fused-silica waveguides. The aims of Mark's work are to determine the relative effectiveness of two different forms of waveguide and to identify the end-facet preparation necessary to minimise unwanted effects like focal ratio degradation.

Petchara Pattarakijwanich is a student from Thailand in his third year at the University of Oxford. Petchara is working with Matthew Colless, Heath Jones, and Chris Springob on the 6dF Galaxy Survey. Within a broad ranging project, Petchara is studying the relationship between the stellar and dynamical masses of 6dFGS with regard to their local environment. Petchara is currently studying physics and will major in astrophysics and condensed matter during his fourth year. He joined the AAO in mid-July and will stay until his studies recommence in October.

Talini Jayawardena is studying for an MEng at the University of Bath, UK. She completed her fourth year of study in May 2009 and will finish the course in June 2010. She is working with Michael Goodwin on the development of a closed loop control system for the 'Starbugs' robotic fibre positioning system, which may be used in future telescopes. The prototype currently under development consists of a camera providing visual data that can be processed to obtain the position of a Starbug to micrometer precision. This information can be used to substantially increase the accuracy of the fibre placement.

## LETTER FROM COONABARABRAN
Rhonda Martin

It is quite some time since the last newsletter but in that time we have had a celebration, a farewell or two and an accident with, fortunately, nobody hurt.

While contractors were using a crane to lift a cooling tower for the new air conditioning unit, it fell over – the crane that is – taking out a 22,000v power line on the way and blacking out Timor Road. It didn't do the cooling tower a lot of good either. Seeing the crane driver managing to climb out through the door and not obviously hurt just made for an exciting lunchtime, however he went into shock which necessitated a trip to hospital, but all was well. Country Energy were quickly on the scene to make the necessary repairs and Timor Road was back up again. There is nothing dull on the mountain!

After 35 years at the AAO a celebration lunch was held for Bob Dean at the AAT with a video link to Epping. Bob is held in high esteem as was evidenced by the packed condition of the lunch room. The food was great, as was the company. Bob has recently been made Telescope Systems Manager which should keep him hopping.

Martin Oestreich, on his way to a new position in Chile, after being Head of Electronics at the AAT, was farewelled at the same time.

The Operations Manager, Chris McCowage, had decided to call it a day and retire so another very satisfying lunch was in order. Have you noticed that everything seems to happen at lunchtime up here? It was a good farewell with old employees come to wish Chris well. Staff gave him the barometer he requested and a beautiful sundial from the skilful hands of Mick Kanonczuk. Epping added to the haul with a very nice pair of binoculars. Chris has been in Europe for the past couple of months, learning how to relax – we wish him all the very best for his retirement.

Doug Gray (who considers that driving to Dubbo is a major undertaking – you can tell he is from a small island!) is our new Operations Manager and is settling in well.

It was a big affair when the ANU's Skymapper was officially opened by the Governor of NSW, Marie Bashir. A wonderful asset for Siding Spring, and for Coonabarabran.

**EPPING NEWS**
Sandra Ricketts

Since the last newsletter we have welcomed yet more new arrivals to the AAO. Michael Goodwin, Sarah Brough and Stuart Barnes introduce themselves below, but they are not the only new members of staff at Epping.

Jeroen Heijmans has joined the instrumentation group as a project engineer, Anthony Heng is a project manager for HERMES, while Ian Saunders is a project manager for the instrument science group and will be working on various projects.

And congratulations to Heath Jones and his wife on the arrival of Evan Jones in April.

Michael Goodwin: I commenced as an instrument scientist with the AAO in June 2009 working under the Instrument Science Group headed by Roger Haynes. Two challenging projects that I am currently assigned to include the Starbugs project and the HERMES data simulator project.

I recently completed my PhD at The Australian National University in 2009 under the supervision of Charles Jenkins. My work involved the observations and modelling of the atmospheric turbulence profile above Siding Spring and Las Campanas Observatories. Using the derived atmospheric turbulence models that characterise each astronomical site I was able to perform adaptive optics simulations to assess the performance of several adaptive optics techniques.

During 2008/2009 I was involved in the monitoring and verification of $CO_2$ measurements from the $CO_2$ Geosequestration demonstration project in Victoria coordinated by the Cooperative Research Centre for Greenhouse Gas Technologies in Canberra. The outcomes of my research included mathematical algorithms (process monitoring and control) to verify $CO_2$ measurements using ecological parameters (e.g., temperature, pressure, humidity, wind speed). I also developed an interactive web chart for the public to view the $CO_2$ measurements.

Since my undergraduate days I have maintained a high interest in astronomical knowledge which was the motivation for pursing a PhD in astronomical instrumentation. As an instrument scientist I look forward to contributing my skills and knowledge to uphold the excellence of the instrumentation group.

Sarah Brough: I joined the AAO at the beginning of July 2009 to spend the next 3 years as an AAO Research Fellow, working with the Head of AAT Science, Dr Andrew Hopkins. I did not have to move far as I have been working at Swinburne University in Melbourne for the last 5 years, following the completion of my PhD at Liverpool John Moores University, UK, in 2004.

My research is mainly concerned with how galaxies evolve and how that depends on their environment – particularly for the most massive ones: Brightest Cluster Galaxies (BCGs). These massive elliptical galaxies are found at the centres of groups and clusters of galaxies and understanding their evolution will help us to understand how both galaxies and the systems they live in evolve.

At the AAO I will continue my involvement with the WiggleZ survey being undertaken at the AAT. This survey will provide the first high-precision measurement of "dark energy" independent of previous supernovae determinations. I will also expand on my current BCG research, using integral field spectrograph observations to characterise the 3D properties of these massive galaxies. I will also be initiating projects with the GAMA (Galaxy And Mass Assembly) survey being undertaken at the AAT, which aims to study structure in the Universe on scales of 1 kpc to 1 Mpc. These observations will add to our understanding of how baryons evolve in the framework of a $\Lambda$CDM standard cosmology.

Stuart Barnes: I started at the AAO in June, 2009 as an Instrument Scientist. I completed my PhD in astronomy at the University of Canterbury, New Zealand in 2004. During my doctoral studies I worked on the design and construction of high resolution spectrographs; namely HERCULES (at Mt John Observatory) and SALT HRS (currently under construction at the University of Durham).

Subsequent to graduation I took first a postdoc and then a research associate position at McDonald Observatory, University of Texas at Austin. During the 3 1/2 years I spent in Texas I worked on a number of instrumentation projects, including design studies for the high resolution optical spectrograph for the Giant Magellan Telescope, as well as the IGRINS instrument (an H and K band high resolution spectrograph for a 4-metre class telescope).

Apart from instrumentation and optical design, I have an interest in the detection of planets via precision radial velocities, and also in the design of low cost spectrographs suited to this purpose.

I am glad to be part of an institution with a proven history in the deployment of successful instruments and I look forward to contributing to the development of future instrumentation at the AAO.

# Chris McCowage declares his innings closed on 29 (years)!

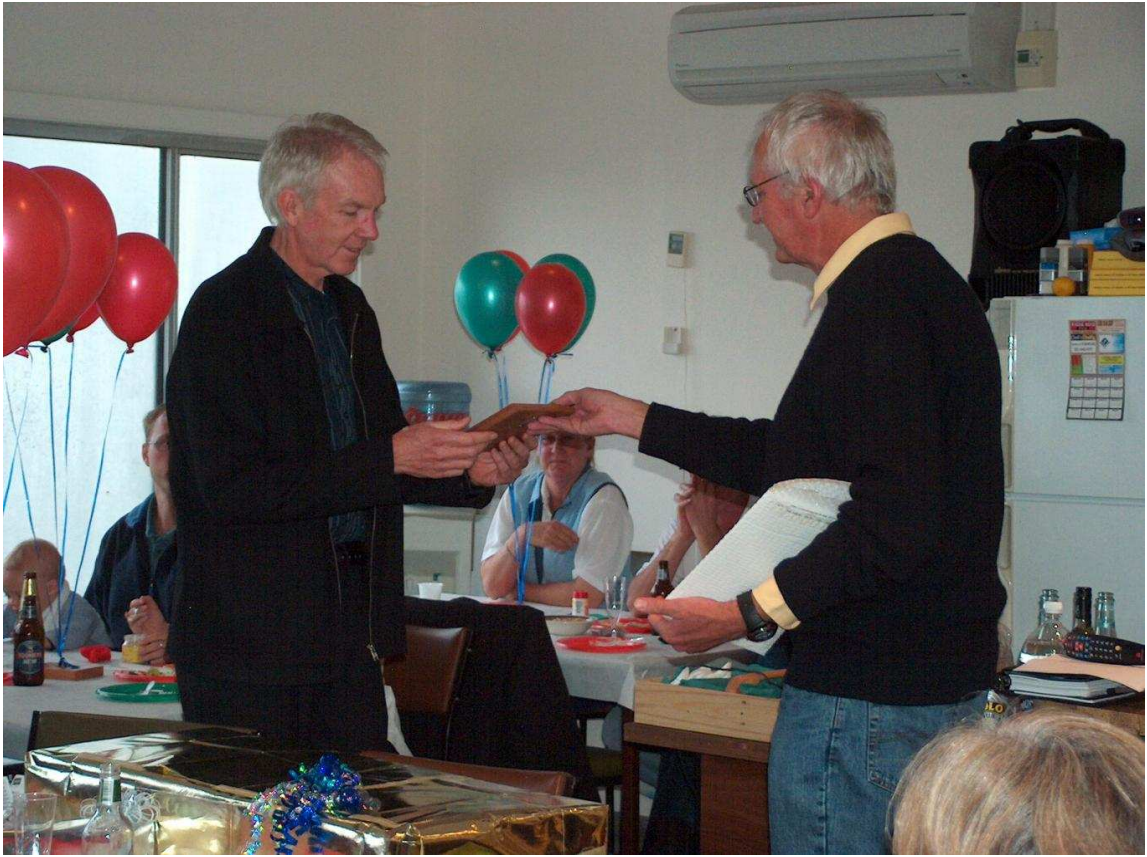

A party was held at the AAT in May to mark the retirement of long-serving Operations Manager Chris McCowage (photo courtesy of Tim Connors).

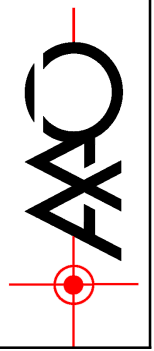